\documentclass[a4paper,fleqn,usenatbib,useAMS]{mnras}

\usepackage{graphicx}	
\usepackage{amsmath}	
\usepackage{amssymb}	
\usepackage{multicol}        
\usepackage{bm}		
\usepackage{pdflscape}	
\usepackage{amsmath,breqn} 

\usepackage[T1]{fontenc}
\usepackage{aecompl}

\title[\mbox{X-ray} flux variability of AGN]{\mbox{X-ray} flux variability of active galactic nuclei observed using NuSTAR}

\author[Priyanka Rani et al.]{
Priyanka Rani,\thanks{E-mail: priyanka@iiap.res.in}
C. S. Stalin and
Suvendu Rakshit
\\
Indian Institute of Astrophysics, Koramangala, Bangalore 560034, India\\
}

\date{Accepted 2016 December 08. Received 2016 December 08; in original form 2016 September 26}

\begin{document}
\label{firstpage}
\pagerange{\pageref{firstpage}--\pageref{lastpage}}
\maketitle

\begin{abstract}

We present results on a systematic study of flux
variability on hourly time-scales
in a large sample of active galactic nuclei (AGN) in the
3$-$79 keV band using data from
Nuclear Spectroscopic Telescope Array. Our sample consists of 4 BL Lac objects (BL Lacs),
3 flat spectrum radio quasars (FSRQs) 24 Seyfert 1, 42 Seyfert
2 and 8 narrow line Seyfert 1 (NLSy1) galaxies.
We find that in the 3$-$79 keV band, about 65\% of the sources in our sample
show significant variations on
hourly time scales. Using Mann-Whitney U-test and Kolmogorov-Smirnov
test, we find no difference
in the variability behaviour between Seyfert 1 and 2 galaxies.
The blazar sources (FSRQs and BL Lacs) in our sample, 
 are 
more variable than Seyfert galaxies
that include Seyfert 1 and Seyfert 2 in the soft (3$-$10 keV),
hard (10$-$79 keV) and total (3$-$79 keV) bands.
NLSy1 galaxies show the highest duty cycle of
variability (87\%), followed by  BL Lacs (82\%), Seyfert galaxies (56\%)
and FSRQs (23\%). We obtained flux doubling/halving time in the hard
\mbox{X-ray} band less than 10 min in 11 sources.
 The flux variations between the hard and soft bands in all
the sources in our sample are consistent with zero lag.

\end{abstract}


\section{Introduction}\label{sec:intro}

There is convincing evidence that accretion onto super-massive black holes 
powers active galactic nuclei (AGN; \citealt{1984ARA&A..22..471R}). A vast 
majority of about 85\% of them emit little or no radio emission, 
these AGN are termed radio-quiet AGN and a minority
of about 15\% called radio-loud AGN have large scale relativistic jets and 
emit copiously in the radio band. It is still not known what triggers 
relativistic jets in only a small fraction of AGN and thereby giving rise to 
the apparent radio-loud radio-quiet dichotomy \citep{2002AJ....124.2364I,2003MNRAS.346..447C}. Both radio-loud and radio-quiet
AGN may be categorized into several sub-classes
such as Seyfert galaxies (Seyfert 1 and 2), radio galaxies, blazars 
comprising flat spectrum radio quasars (FSRQs) and BL Lac objects 
(BL Lacs) etc., According to the 
unification model, the observed differences between the different types
of AGN are in part due to orientation effects \citep{1993ARA&A..31..473A,
1995PASP..107..803U}. One of the defining characteristics of AGN is that
they show flux variations across the entire accessible electromagnetic 
spectrum \citep{1995ARA&A..33..163W,1997ARA&A..35..445U} over a wide range of 
timescales ranging from minutes to 
months and years. 
Though flux variations in AGN are known since their discovery,
we still do not completely understand the physical processes that cause 
such variations \citep{2006ASPC..360...85M}. Also, it is not unambiguously 
known 
if the same physical processes are responsible for the observed
flux variations in different types of AGN. In spite of our lack of a clear knowledge
on the cause of flux variations, this characteristic of AGN
when probed particularly in the \mbox{X-ray} band can give important constraints
on the physical properties in the innermost regions of AGN and can even provide
clues on the observed radio-loud and radio-quiet dichotomy  
\citep{2002A&A...391..875G}. 

Among the \mbox{X-ray} band, hard \mbox{X-ray} (with energies 
greater than 15 keV) variability in particular can provide
clues about the physics of the central regions of  
AGN and kinematics of the jet as it is less affected by 
absorption than soft X-rays, when the line of sight hydrogen column density is 
less than 10$^{23}$ cm$^{-2}$ \citep{2014A&A...563A..57S}. Hard \mbox{X-rays}
can thus serve as an effective wavelength range to probe the intrinsic
properties of the different classes of AGN. Therefore, the  study of hard 
\mbox{X-ray} properties of the various types of AGN in general and hard 
\mbox{X-ray}
flux variability in particular can be used to test the validity of the 
unification model of AGN \citep{1994MNRAS.267..743G, 
1995PASP..107..803U,2012agn..book.....B}. 
Various models are
available in the literature on the physical processes that cause hard 
\mbox{X-ray} 
emission in AGN. According to \cite{1993ApJ...413..507H}, the hard 
\mbox{X-ray}
emission in radio-quiet AGN is due to the Comptonization of the soft
accretion disk photons by a plasma of hot electrons situated above the disk.
In the case of radio-loud AGN, in addition to the process that produces
hard \mbox{X-ray} emission in radio-quiet AGN, there are additional 
contributions through inverse Compton (IC) processes from
relativistic electrons in the jet. 

In any given X-ray band, the observed characteristics of different types 
of AGN depend on the physical processes that contribute to
the observed spectral energy distribution (SED) in that band. In blazars, the 
broad band SED shows two broad emission peaks: the low energy peak is produced
by synchrotron emission and the high energy peak is produced by 
IC emission process. The seed photons for IC can be from
various sources such as the synchrotron photons themselves 
\citep{1981ApJ...243..700K}, 
accretion disk 
\citep{1993ApJ...416..458D,1997A&A...324..395B}, the 
broad line region (BLR, \citealt{1994ApJ...421..153S,1996MNRAS.280...67G}) or 
the torus \citep{2000ApJ...545..107B}. These external photon fields
contribute differently in individual blazars and this can explain the 
varied nature of the high energy component in the SED of blazars. Depending on 
the location of the synchrotron peak in their SEDs, blazars are further 
subdivided 
\citep{1995ApJ...444..567P,2010ApJ...716...30A} into high synchroton peaked
blazars (HSP, with synchrotron emission peaking at X-ray energies with 
$\nu_s > 10^{15}$ Hz), intermediate synchrotron peaked
blazars (ISP, 10$^{14}$ $<$   $\nu_s < 10^{15}$ Hz and low synchrotron peaked 
blazars (LSP, with synchrotron emission peaking in the IR; $\nu_s < 10^{14}$ 
Hz). In HSPs, the X-ray spectrum
falls in the synchrotron region, thereby showing large amplitude X-ray 
variations at long as well as short time scales 
\citep{1993ApJ...404..112S,2000ApJ...543..124T,
2002ApJ...572..762Z,2005ApJ...629..686Z}, however, less variable in
the optical \citep{1996ASPC..110...64H}. On the other hand, LSPs are more 
variable in the optical \citep{1996ASPC..110...64H} compared
to the X-ray \citep{2016MNRAS.462.1508G} band which falls in the IC region of 
their SED. Nonetheless, it is not always the case and there are instances
where the X-ray spectrum is found to have contribution both from the 
high energy tail of the synchrotron component and the IC component. This 
has been noticed in several LSP blazars
\citep{2000A&A...354..431T,2003ApJ...584..153T,2016MNRAS.458...56W} and in 
one HSP blazar \citep{2016ApJ...827...55K} namely Mrk 421.

Blazars are known to show fast variability. Detection of such fast
variability time scales can set an upper limit on the size of the 
emission region as $R_s$ $<$ c $\delta$ t$_{var}$, via light travel time arguments. 
Here, $\delta$ is the Doppler factor, and t$_{var}$ is the variability
time scale. Such
fast time scale of variations (in the order of minutes) characterised by the 
flux doubling/halving time scale  are often seen in high energy 
$\gamma$-rays \citep{1996Natur.383..319G,2007ApJ...664L..71A}. Also, in HSP 
blazars correlated X-ray and $\gamma$-ray 
variations are found \citep{2015A&A...578A..22A,2016ApJ...819..156B} which is very well expected in the one zone
leptoinc model of blazar emission, wherein, both hard X-rays and
$\gamma$-rays are produced by the same population of relativistic 
electrons in the jet. Detection of such  short time scale X-ray flares with/without a
high energy $\gamma$-ray counterpart will constrain the radiative processes
operating in the sub-parsec scales of AGN. 
Detection of very small flux doubling/halving
time scale is thus an important addition to the knowledge of AGN flux 
variability. Previous efforts to search for flux doubling/halving time scale
$<$ 15 minutes in the X-ray band were negative \citep{2015ApJ...802...33P}.
Therefore, any evidences for the presence of minute scale flux 
doubling/halving time scale is more important.

In the soft \mbox{X-ray} band, with \mbox{X-rays} having energies $<$ 10 keV, numerous 
results on the flux variability nature of 
AGN on time scales ranging from hours to months and years are available
mainly based on observations from
{\it Rossi X-ray Timing Explorer} (RXTE) and XMM-Newton \citep{1997ApJ...476...70N,1998ApJ...503..607F,
1999ApJ...524..667T,2002MNRAS.332..231U,
2003ApJ...598..935M,2008A&A...486..411S,2010LNP...794..203M}.  On the other
hand, studies on the hard \mbox{X-ray} variability of AGN are very limited. 
They include hard \mbox{X-ray} spectral variability based on observations
from BeppoSAX \citep{2000ApJ...540..131P}, INTEGRAL \citep{2013A&A...549A..73P}
and Suzaku \citep{2012ApJ...745...93R}. Long term hard X-ray monitoring
observations have been recently possible owing to the observations
by the Burst Alert Telescope (BAT; \citealt{2005SSRv..120..143B}) instrument 
on board the {\it Swift} satellite \citep{2004ApJ...611.1005G}. Because of 
{\it Swift}/BAT's observing capability in survey mode and its large field of
view of $\sim$1.4 sr, we have long term monitoring data on a large sample of 
bright AGN from BAT. Using this data set, AGN have 
been studied for long term variability on time scale of days to years  
\citep{2004ApJ...611.1005G,2012A&A...537A..87C, 2014A&A...563A..57S}.
However, studies on the hard X-ray variations on time scales of the order
of hours are available only on few individual objects
such as Mrk 421  
\citep{2015ApJ...811..143P} using the focusing 
hard X-ray telescope {\it Nuclear Spectroscopic Telescope Array}
({\it NuSTAR}; \citealt{2013ApJ...770..103H}), 
sources such as BL Lac \citep{2003A&A...408..479R}, 
ON 231 \citep{2000A&A...354..431T}, Mrk 421 \citep{1999ApJ...526L..81M}, 
Mrk 501 \citep{1998ApJ...492L..17P} etc. from BeppoSAX as well as 
observations from RXTE \citep{2008ApJ...677..906F}.
These observations pertaining to sources in different brightness states 
indicate different variability behaviour between hard and soft bands, which
point to different physical processes contributing to the soft and hard
X-ray emission.

Observations using NuSTAR with its high sensitivity and wide spectral 
coverage between 3 $-$ 79 keV is ideal to constrain the X-ray emission 
processes that contribute to the X-ray emission over a wide spectral
range in AGN. We  
investigate here for the first time the hard X-ray 
flux variations on hourly time scales in the largest sample of AGN
that includes both radio-loud and radio-quiet sources. The main motivation here
is to understand (i) the similarities and differences in the flux variability 
nature of different types of AGN over a wide waveband of
3 $-$ 79 keV, (ii) the differences if any between flux variations in soft
and hard X-ray bands and (iii) the presence of fast variations with
flux doubling/halving time scale of few minutes. For this we use the publicly 
available data from observations by {\it NuSTAR}. 
The structure of the paper is as follows. In section 2, we describe the 
sample used and data reduction. In Section 3
we describe the various analysis carried out on the data set. We discuss
our results in Section 4 and summarize the results in Section 5.
We adopt a cosmology with  $H_0=71$~km~s$^{-1}$~Mpc$^{-1}$, 
$\Omega_\Lambda = 0.73$ and $\Omega_m = 0.27$. 

\begin{figure}
\begin{center}
\includegraphics[width=7cm,height=6cm]{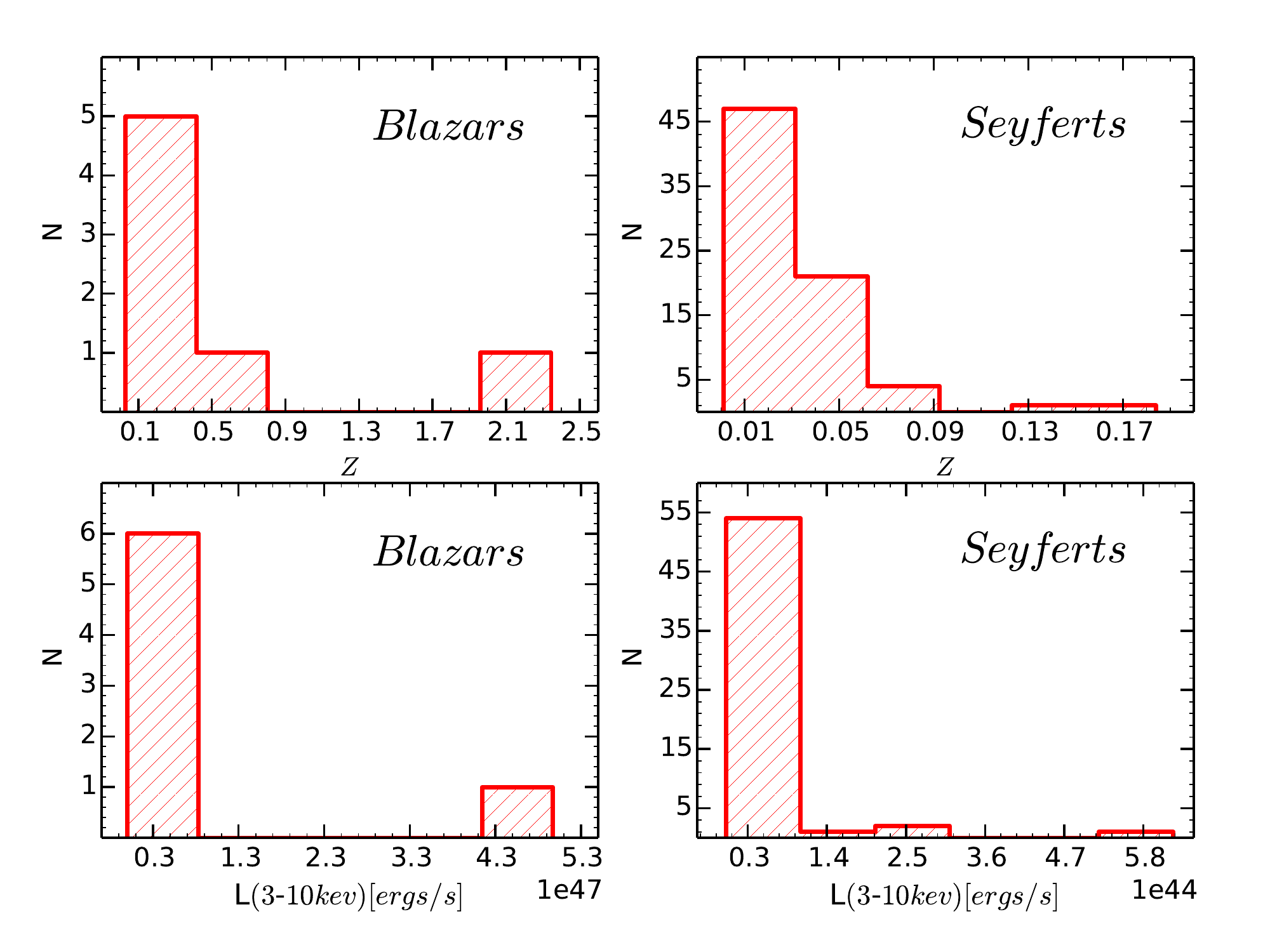}
\caption{\label{distribution}The distribution of redshift and the 3$-$10 keV luminosity of 
the sources. The left hand panels are for blazars, that
includes FSRQs and BL Lacs, and the right hand panel are for
Seyferts comprising Seyfert 1 and 2 galaxies.}
\end{center}
\end{figure}

\section{Sample and  Data Reduction}

The goal of this work is to find the hard X-ray flux variations in different
types of AGN. For this we have used the data from  
{\it NuSTAR}. {\it NuSTAR} (\citealt{2013ApJ...770..103H})  
launched in June 2012, is 
the first focusing hard X-ray telescope.  It has 
a field of view 13$'$ x 13$'$. It consists of two co-aligned X-ray detector 
pairs with corresponding focal plane modules FPMA and FPMB\footnote{https://heasarc.gsfc.nasa.gov/docs/nustar/}.
{\it NuSTAR} observations of all AGN that have become public before
March 2015 have been used in this work. Our sample thus consists of 81 AGN, 
among these 4 are BL Lacs, 3 are FSRQs, 24 are 
Seyfert 1 galaxies, 42 are Seyfert 2 galaxies and 8 are narrow line
Seyfert 1 (NLSy1) galaxies. The details of the sources
used in this study are given in Table ~\ref{sourcetable}.  The distribution of 
redshifts and the 3$-$10 keV luminosity of the sources studied are shown in 
Figure~\ref{distribution} separately for radio-quiet sources and blazars.

\begin{figure}
\hspace*{-0.2cm}\includegraphics[width=10cm,height=8cm]{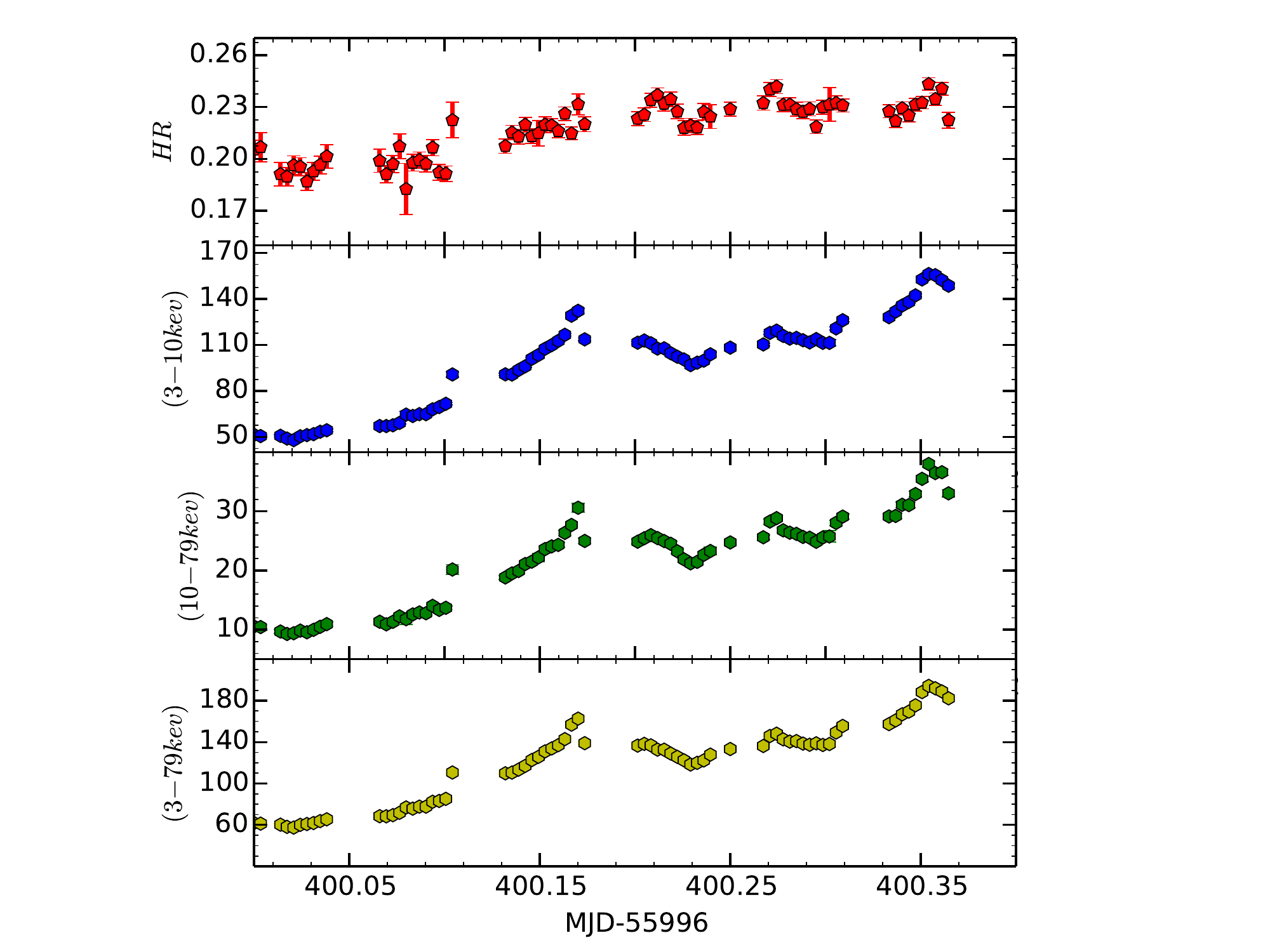}
\caption{\label{lcmrk421}Light curves of the BL Lac object  Mrk 421 corresponding to the 
observational ID 60002023031 and observed on 2013-04-14 for a duration of 15606 sec. From the 
top to the bottom
are shown the HR variation, flux variations in the energy ranges
of 3$-$10 keV (soft band), 10$-$79 keV (hard band) and 
3$-$79 keV (total band) respectively.
Each point corresponds to a binning of 300 seconds
and the fluxes are in units of counts s$^{-1}$.}
\end{figure}

Reduction of the NuSTAR data was done using the data 
analysis software {\it \small{NuSTARDAS}} v.1.4.1 distributed by the High Energy Astrophysics
Archive Research Center (HEASARC). The calibrated, cleaned and screened event 
files were generated using the {\it nupipeline} task and using CALDB 20141107. A 
circular region of $60''$ radius was taken to extract the source and background 
counts on the same detector. We extracted the 5 min binned light curves in 
the energy range of 3$-$79 keV in each focal plane module FPMA and FPMB using 
the {\it nuproducts} package available in {\it \small{NuSTARDAS}}. Further, these 
light curves were divided into two bands: 3$-$10 keV and 10$-$79 keV. To generate 
light curves, the count rates of the two modules FPMA and FPMB are combined 
using `lcmath' task included in FTOOLS V.4.0. Sample
light curves of the BL Lac object Mrk 421 are shown
in Figure~\ref{lcmrk421}. 

\begin{table*}
\caption{\label{sourcetable}Information on the sources studied for flux variability. The quoted magnitudes are the V-band magnitudes taken from \citet{2010A&A...518A..10V} except for those appended with a * for which it is from NED (\url{http://ned.ipac.caltech.edu/}).}
\footnotesize
\resizebox{\linewidth}{!}{%
\begin{tabular}{llllllllllll} \hline
Name  &  $\alpha$(2000)   &  $\delta$(2000)  & V  & $z$   &   Type  &   Name  &  $\alpha$(2000)   &  $\delta$(2000)  & V  & $z$   &   Type\\ 
 &    &   & (mag)  &   &    &    &    &   & (mag)  &   &   \\ \hline
Mrk 335               &  00:06:19.5  &  $+$20:12:11  &   13.85   &  0.026    &  Sy1   & 3C 273                 &  12:29:06.7  &  $+$02:03:08  &   12.85   &  0.158    &  FSRQ   \\  
NGC 424               &  01:11:27.7  &  $-$38:05:10  &   14.12   &  0.011    &  Sy2   & 3C 279                 &  12:56:11.1  &  $-$05:47:22  &   17.75   &  0.538    &  FSRQ    \\
NGC 513               &  01:24:26.8  &  $+$33:47:58  &   13.40   &  0.019    &  Sy2   & Mrk 231                &  12:56:14.2  &  $+$56:52:25  &   13.84   &  0.041    &  Sy1    \\  
NGC 612               &  01:33:57.8  &  $-$36:29:36  &   13.20   &  0.030    &  Sy2   & Mrk 248                &  13:15:17.2  &  $+$44:24:26  &   15.10   &  0.036    &  Sy2    \\  
MCG $+$01$-$05$-$047  &  01:52:49.0  &  $-$03:26:49  &   14.24   &  0.017    &  Sy2   & MCG $-$06$-$30$-$15    &  13:35:53.4  &  $-$34:17:48  &   13.61   &  0.008    &  Sy1    \\  
NGC 788               &  02:01:06.5  &  $-$06:48:56  &   12.76   &  0.013    &  Sy2   & NGC 5252               &  13:38:15.9  &  $+$04:32:33  &   14.21   &  0.022    &  Sy2    \\  
NGC 985               &  02:34:37.8  &  $-$08:47:15  &   14.28   &  0.043    &  Sy1   & NGC 5273               &  13:42:08.3  &  $+$35:39:15  &   13.12   &  0.003    &  Sy2     \\  
NGC 1068              &  02:42:40.7  &  $-$00:00:47  &   10.83   &  0.003    &  Sy2   & Mrk 273                &  13:44:42.1  &  $+$55:53:13  &   14.91   &  0.037    &  Sy2     \\  
1H 0323$+$342         &  03:24:41.2  &  $+$34:10:45  &   15.72   &  0.061    &  NLSy1 & IC 4329A                &  13:49:19.3  &  $-$30:18:34  &   13.66   &  0.016    &  Sy1     \\  
NGC 1320              &  03:24:48.7  &  $-$03:02:32  &   14.00   &  0.009    &  Sy2   & PKS 1409$-$651         &  14:13:09.8  &  $-$65:20:17  &   12.10   &  0.001    &  Sy2     \\
NGC 1365              &  03:33:36.4  &  $-$36:08:24  &   12.95   &  0.006    &  Sy2   & NGC 5506               &  14:13:14.8  &  $-$03:12:26  &   14.38   &  0.007    &  Sy1     \\  
3C 120                &  04:33:11.1  &  $+$05:21:15  &   15.05   &  0.033    &  Sy1   & NGC 5548               &  14:17:59.6  &  $+$25:08:13  &   13.73   &  0.017    &  Sy1     \\ 
MCG $+$03$-$13$-$01   &  04:46:29.7  &  $+$18:27:40  &   15.00   &  0.016    &  Sy2   & NGC 5674               &  14:33:52.3  &  $+$05:27:30  &   13.70   &  0.025    &  Sy2     \\    
IRAS 04507$+$0358     &  04:53:25.7  &  $+$04:03:42  &   15.00   &  0.030    &  Sy2   & Mrk 477                &  14:40:38.1  &  $+$53:30:16  &   15.03   &  0.038    &  Sy2     \\    
XSS J05054$-$2348     &  05:05:45.7  &  $-$23:51:14  &   17.00   &  0.035    &  Sy2   &  NGC 5728              &  14:42:23.9  &  $-$17:15:11  &   13.40   &  0.009    &  Sy2     \\    
ZW 468.002  	      &  05:08:19.7  &  $+$17:21:47  &   13.50   &  0.017    &  Sy2   & IGR J14552$-$5133       &  14:55:17.8  &  $-$51:34:17  &   17.10    &  0.016    &  NLSy1   \\   
ARK 120               &  05:16:11.4  &  $-$00:09:00  &   13.92   &  0.033    &  Sy1   &  IC 4518A                &  14:57:41.2  &  $-$43:07:56  &   15.00     &  0.016    &  Sy2     \\   
IRAS 05189$-$2524     &  05:21:01.4  &  $-$25:21:45  &   14.75   &  0.042    &  Sy2   &  SWIFT J1514.5-8123 &  15:14:42.0  &  $-$81:23:38  &   $*$17.3   &  0.068    &  Sy1     \\   
NGC 2110              &  05:52:11.4  &  $-$07:27:23  &   13.51   &  0.007    &  Sy2   &  Mrk 290                   &  15:35:52.3  &  $+$57:54:09  &   15.30   &  0.030    &  Sy1       \\  
NGC 2273              &  06:50:08.7  &  $+$60:50:45  &   13.54   &  0.006    &  Sy2   &  Mrk 501                   &  16:53:52.2  &  $+$39:45:36  &   13.78   &  0.033    &  BL Lac    \\   
1H 0707$-$495         &  07:08:41.5  &  $-$49:33:06  &   15.70   &  0.041    &  NLSy1 &  MCG $+$05$-$40$-$026      &  17:01:07.8  &  $+$29:24:25  &   15.78   &  0.036    &  NLSy1     \\  

IRAS 07245$-$3548     &  07:26:26.3  &  $-$35:54:22  &   16.80   &  0.029    &  Sy2   &  NGC 6300                  &  17:16:59.2  &  $-$62:49:05  &   13.08   &  0.003    &  Sy2     \\
Mrk 9                  &  07:36:57.0  &  $+$58:46:13  &   14.37   &  0.039    &  Sy1   &  PDS 456                   &  17:28:19.9  &  $-$14:15:56  &   14.03   &  0.184    &  NLSy1     \\  
IRAS 07378$-$3136     &  07:39:44.7  &  $-$31:43:02  &   $*$16.8   &  0.025    &  Sy2 &  IGR J18244$-$5622         &  18:24:19.4  &  $-$56:22:09  &   14.40  &  0.017    &  Sy2        \\  
Mrk 1210                 &  08:04:05.9  &  $+$05:06:50  &   13.70   &  0.013    &  Sy2 & LEDA 3097193               &  18:26:32.4  &  $+$32:51:30  &   $....$    &  0.022    &  Sy2     \\    
FAIRALL 0272              &  08:23:01.1  &  $-$04:56:05  &   16.00   &  0.021    &  Sy2 & 3C 382                   &  18:35:03.4  &  $+$32:41:47  &   15.39   &  0.058    &  Sy1       \\  
FAIRALL 1146              &  08:38:30.8  &  $-$35:59:33  &   16.10   &  0.031    &  Sy1 & H 1834$-$653               &  18:38:20.5  &  $-$65:25:39  &   14.53   &  0.013    &  Sy2       \\ 
SWIFT J0845.0$-$3531  &  08:45:21.4  &  $-$35:30:24  &   $....$    &  0.137    &  Sy1 & 3C 390.3                    &  18:42:09.0  &  $+$79:46:17  &   15.38   &  0.057    &  Sy1        \\ 
MCG $+$01$-$24$-$012     &  09:20:46.2  &  $-$08:03:21  &   13.70   &  0.020    &  Sy2   & 2E 1849.2$-$7832           &  18:57:07.7  &  $-$78:28:21  &  $*$14.5    &  0.042    &  Sy1     \\ 
MCG $+$04$-$22$-$042     &  09:23:43.1  &  $+$22:54:33  &   14.80   &  0.033    &  NLSy1 & IGR J19473$+$4452    	   &  19:47:19.4  &  $+$44:49:42  &   15.70   &  0.053    &  Sy2  \\ 
MCG $-$05$-$23$-$16      &  09:47:40.2  &  $-$30:56:54  &   13.69   &  0.008    &  Sy1   & 3C 403                     &  19:52:15.9  &  $+$02:30:24  &   16.50   &  0.059    &  Sy2       \\ 
3C 227                   &  09:47:45.1  &  $+$07:25:20  &   16.97   &  0.086    &  Sy1  & MCG $+$07$-$41$-$003       &  19:59:28.3  &  $+$40:44:02  &   15.10   &  0.056    &  Sy2       \\ 
NGC 3079                 &  10:01:58.5  &  $+$55:40:50  &   12.18   &  0.004    &  Sy2  & IGR J20187$+$4041          &  20:18:38.7  &  $+$40:41:00  &   $....$    &  0.014    &  Sy2     \\
Mrk 728                  &  11:01:01.8  &  $+$11:02:50  &   16.93   &  0.036    &  Sy2  & IC 5063                     &  20:52:02.2  &  $-$57:04:08  &   13.60   &  0.011    &  Sy2       \\ 
Mrk 421                  &  11:04:27.2  &  $+$38:12:32  &   12.90   &  0.031    &  BL Lac& IGR J21277$+$5656          &  21:27:44.9  &  $+$56:56:40  &   18.79    &  0.014    &  NLSy1    \\ 
NGC 3516                 &  11:06:47.4  &  $+$72:34:07  &   12.40   &  0.009    &  Sy1  & IRAS F21318$-$2739        &  21:34:45.1  &  $-$27:25:55  &  $*$16.36   &  0.067    &  Sy1     \\
Mrk 732                  &  11:13:49.8  &  $+$09:35:10  &   14.17   &  0.030    &  Sy1  & PKS 2149$-$306             &  21:51:55.4  &  $-$30:27:54  &   17.90   &  2.345    &  FSRQ      \\  
NGC 4051                 &  12:03:09.6  &  $+$44:31:53  &   12.92   &  0.002    &  NLSy1& PKS 2155$-$304             &  21:58:52.0  &  $-$30:13:32  &   13.09   &  0.116    &  BL Lac    \\  
NGC 4151                 &  12:10:32.5  &  $+$39:24:21  &   11.85   &  0.003    &  Sy1  & BL Lac                      &  22:02:43.3  &  $+$42:16:39  &   14.72   &  0.069    &  BL Lac    \\   
WAS 49b                   &  12:14:17.8  &  $+$29:31:43  &   15.40   &  0.064    &  Sy2  & NGC 7582                    &  23:18:23.5  &  $-$42:22:14  &   13.57   &  0.005    &  Sy1       \\                              &              &               &           &           &            \\
NGC 4395                 &  12:25:48.9  &  $+$33:32:48  &   10.27   &  0.001    &  Sy2  &                            &              &               &           &           &            \\ \hline
\end{tabular}}
\end{table*}

\begin{figure}
\hspace*{-0.5cm}\includegraphics[width=9cm,height=7cm]{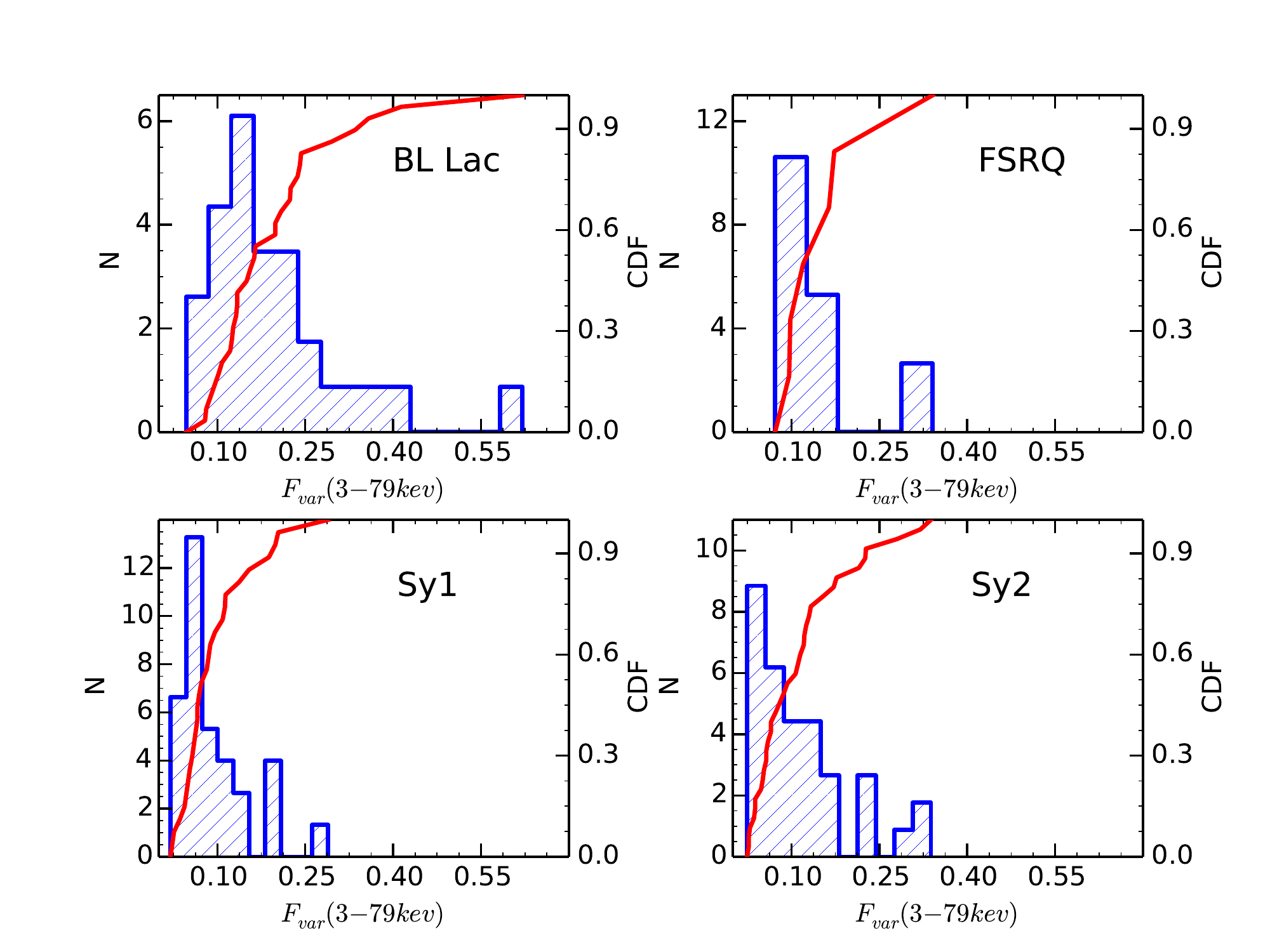}
\caption{\label{flux_distribution}Histogram and cumulative distribution function (CDF) of $F_{\rm{var}}$ in the 3$-$79 keV band for 
different classes of AGN. The values of CDF are given in the right of each figure. }
\end{figure}

\section{Analysis}
\subsection{Variability Amplitude}
To characterize  the flux variations we used the fractional root mean square (rms)
variability amplitude $F_{\rm{var}}$ \citep{2002ApJ...568..610E, 2003MNRAS.345.1271V}. This
gives an estimate of the intrinsic variability amplitude relative to the mean count 
rate exceeding
the measurement noise in 
the light curves. $F_{\rm{var}}$ is defined as
\begin{equation}
\centering
F_{\rm{var}}=\sqrt{{S^{2}-{\bar{\sigma^{2}}_{\rm{err}}}\over\bar{x}^2}}
\end{equation}
where $S^2$ is the sample variance, $\bar{x}$ is the arithmetic mean of $x_i$ 
and $\bar{\sigma^{2}}_{\rm{err}}$ represents the mean square error, given by
\begin{equation}
\centering
S^2={1\over{N-1}}\sum_{i=1}^{N}(x_i-\bar{x})^2
\end{equation}
\begin{equation}
\bar{\sigma^{2}}_{\rm{err}}={1\over{N}}\sum_{i=1}^{N}\sigma^{2}_{\rm{err,i}}
\end{equation}
The uncertainty in $F_{\rm{var}}$ is given by
\begin{equation}
 \centering
          err(F_{\rm{var}})=\sqrt{\Bigg(\sqrt{1\over{2N}}{\bar{{\sigma^{2}}_{\rm{err}}}\over{{\bar{x}}^2}{F_{\rm{var}}}}\Bigg)^2+\Bigg(\sqrt{\bar{\sigma^{2}}_{\rm{err}}\over{N}}{1\over\bar{x}}\Bigg)^2}
\end{equation}
$F_{\rm{var}}$ and $err(F_{\rm{var}})$ were calculated for each binned light 
curve. Variability analysis of the sample
was carried out in soft (3$-$10 keV), hard (10$-$79 keV) and total
(3$-$79 keV) bands. An object is considered variable if $F_{\rm{var}}$
(significant at 1 $\sigma$) 
is greater than zero. Calculated values of $F_{\rm{var}}$ for sources that are found
to be variable  are given in Tables~\ref{fbl}, \ref{ffsrq}
, \ref{fsy1}, \ref{fsy2} and \ref{fnlsy1} for BL Lacs, 
FSRQs, Seyfert 1 galaxies, Seyfert 2 galaxies and NLSy1 galaxies
respectively. About 65\% of the sources in our sample are 
found to be variable.  

\begin{figure*}
\includegraphics[width=11cm,height=9cm]{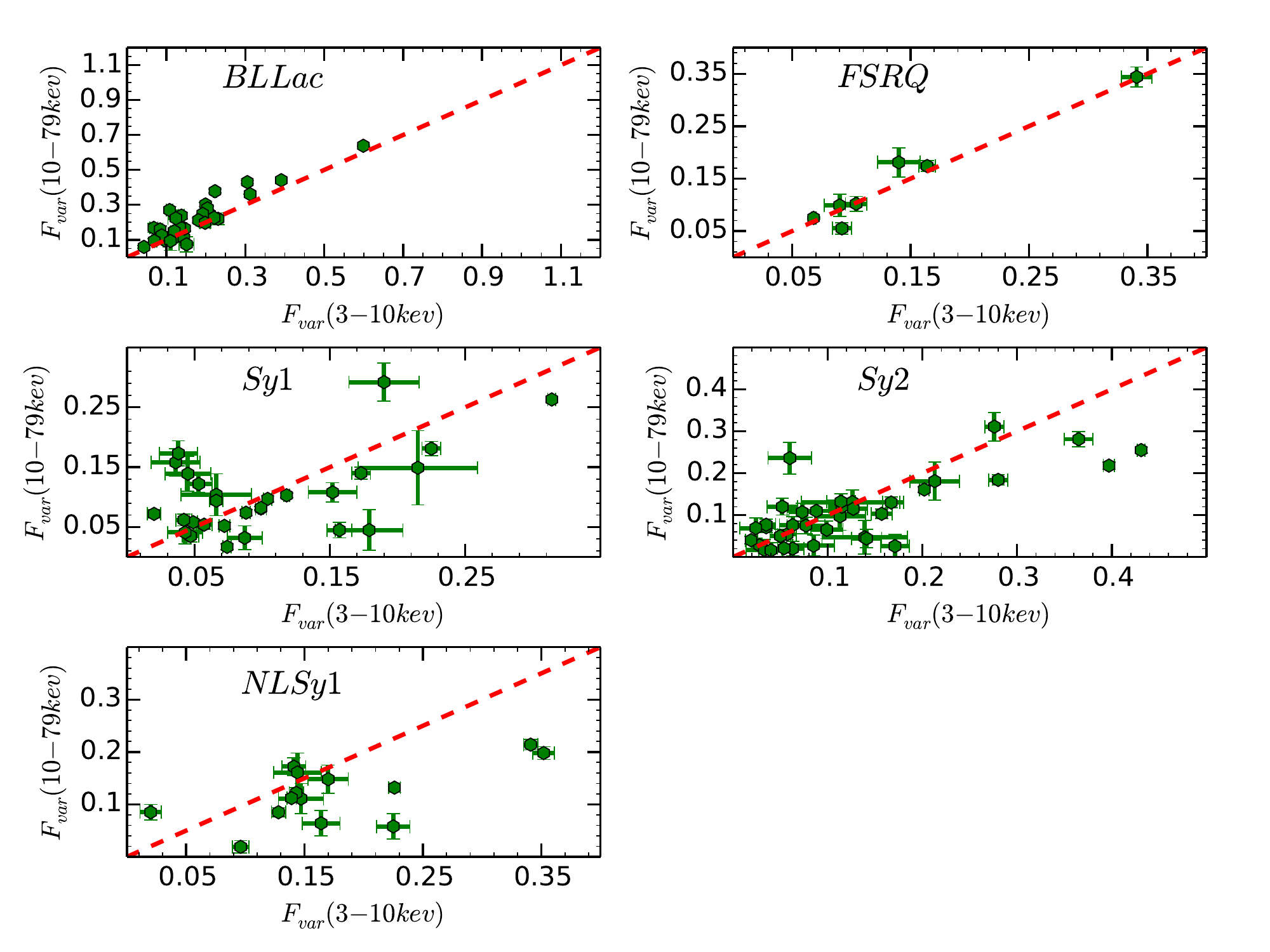}
\caption{\label{correlation}Correlation of flux variations between soft and hard bands for the different classes
of AGN. The dashed lines have a slope of unity and indicate identical variation of $F_{\rm{var}}$ values between
soft and hard bands.}
\end{figure*}

The number of BL Lacs, FSRQs and NLSy1 galaxies studied here is small 
(compared to Seyfert 1 and 2 galaxies) to consider the variability shown
by these objects as a representation of their class as a whole. 
Nevertheless, we calculated the weighted mean variability of the 
different classes of AGN
in different X-ray bands namely soft, hard and total bands and the results are 
presented in 
Table~\ref{afvar}. 
In the total NuSTAR band we find an average $F_{\rm{var}}$ of 0.310 $\pm$ 0.138 
and 0.078 $\pm$ 0.028 for BL Lacs and FSRQs respectively. 
In the radio-quiet category, the 
average $F_{\rm{var}}$ values for Seyfert 1 and Seyfert 2 galaxies are
 0.092 $\pm$ 0.061 and 0.149 $\pm$ 0.108 respectively.
The histogram and 
cumulative distribution of  $F_{\rm{var}}$ in the total band for different classes
of AGN are shown in Figure~\ref{flux_distribution}. 


\begin{table*}
\caption{\label{fbl}Characteristics of flux variations of BL Lac objects}
\centering
\scalebox{1.1}{
\begin{tabular}{@{}llllllll@{}}
\hline
Name  & Type & OBS ID  & Obs. date  &  Exposure & \multicolumn{3}{c}{$F_{\rm{var}} \pm err(F_{\rm{var}})$}  \\
      &     &        &          &  (secs)      &  3$-$10 keV     & 10$-$79 keV     &  3$-$79 keV     \\
\hline
Mrk 421        & BL Lac   & 10002015001 & 2012-07-07 & 42034 & 0.209$\pm$0.003  & 0.247$\pm$0.008  & 0.225$\pm$0.004    \\
                 &         & 10002016001 & 2012-07-08 & 24885 & 0.305$\pm$0.003  & 0.430$\pm$0.009  & 0.358$\pm$0.004    \\
                 &         & 60002023002 & 2013-01-02 & 9152  & 0.068$\pm$0.010  & 0.168$\pm$0.027  & 0.088$\pm$0.013    \\
                 &         & 60002023004 & 2013-01-10 & 22633 & 0.119$\pm$0.008  & 0.146$\pm$0.021  & 0.132$\pm$0.010    \\
                 &         & 60002023006 & 2013-01-15 & 24182 & 0.199$\pm$0.004  & 0.302$\pm$0.012  & 0.243$\pm$0.005    \\
                 &         & 60002023008 & 2013-01-20 & 24968 & 0.084$\pm$0.007  & 0.159$\pm$0.020  & 0.095$\pm$0.009    \\
                 &         & 60002023010 & 2013-02-06 & 19307 & 0.097$\pm$0.005  & 0.095$\pm$0.014  & 0.108$\pm$0.006    \\
                 &         & 60002023012 & 2013-02-12 & 14780 & 0.204$\pm$0.005  & 0.280$\pm$0.012  & 0.241$\pm$0.006    \\
                 &         & 60002023014 & 2013-02-16 & 17359 & 0.231$\pm$0.013  & 0.220$\pm$0.034  & 0.237$\pm$0.020    \\
                 &         & 60002023016 & 2013-03-04 & 17252 & 0.122$\pm$0.005  & 0.115$\pm$0.016  & 0.122$\pm$0.007    \\
                 &         & 60002023018 & 2013-03-11 & 17474 & 0.144$\pm$0.005  & 0.112$\pm$0.014  & 0.125$\pm$0.006    \\
                 &         & 60002023020 & 2013-03-17 & 16558 & 0.088$\pm$0.004  & 0.125$\pm$0.012  & 0.102$\pm$0.006    \\
                 &         & 60002023022 & 2013-04-02 & 24772 & 0.223$\pm$0.004  & 0.378$\pm$0.011  & 0.295$\pm$0.006    \\
                 &         & 60002023024 & 2013-04-10 & 5758  & 0.146$\pm$0.005  & 0.164$\pm$0.013  & 0.165$\pm$0.006    \\
                 &         & 60002023025 & 2013-04-11 & 57509 & 0.599$\pm$0.001  & 0.638$\pm$0.004  & 0.621$\pm$0.002    \\
                 &         & 60002023027 & 2013-04-12 & 7630  & 0.134$\pm$0.002  & 0.172$\pm$0.005  & 0.150$\pm$0.002    \\
                 &         & 60002023029 & 2013-04-13 & 16510 & 0.221$\pm$0.002  & 0.226$\pm$0.005  & 0.224$\pm$0.002    \\
                 &         & 60002023031 & 2013-04-14 & 15606 & 0.312$\pm$0.001  & 0.361$\pm$0.002  & 0.335$\pm$0.001    \\
                 &         & 60002023033 & 2013-04-15 & 17278 & 0.192$\pm$0.006  & 0.248$\pm$0.005  & 0.199$\pm$0.008    \\
                 &         & 60002023035 & 2013-04-16 & 20279 & 0.391$\pm$0.002  & 0.441$\pm$0.004  & 0.414$\pm$0.002    \\
                 &         & 60002023037 & 2013-04-18 & 17795 & 0.181$\pm$0.005  & 0.212$\pm$0.013  & 0.199$\pm$0.006    \\
                 &         & 60002023039 & 2013-04-19 & 15958 & 0.120$\pm$0.005  & 0.149$\pm$0.013  & 0.134$\pm$0.006    \\
Mrk 501          & BL Lac & 60002024004 & 2013-05-08 & 26141 & 0.138$\pm$0.007  & 0.238$\pm$0.019  & 0.163$\pm$0.008    \\
                 &         & 60002024006 & 2013-07-12 & 10857 & 0.043$\pm$0.004  & 0.059$\pm$0.008  & 0.047$\pm$0.005    \\
                 &         & 60002024008 & 2013-07-13 & 10343 & 0.069$\pm$0.007  & 0.094$\pm$0.012  & 0.081$\pm$0.007    \\
PKS 2155$-$304   & BL Lac & 10002010001 & 2012-07-08 & 33838 & 0.108$\pm$0.008  & 0.270$\pm$0.020  & 0.127$\pm$0.010    \\
                 &         & 60002022004 & 2013-07-16 & 13856 & 0.124$\pm$0.020  & 0.223$\pm$0.046  & 0.134$\pm$0.022    \\
                 &         & 60002022008 & 2013-08-08 & 13496 & 0.110$\pm$0.040  & 0.093$\pm$0.054  & 0.156$\pm$0.054    \\
                 &         & 60002022012 & 2013-08-26 & 11356 & 0.198$\pm$0.013  & 0.197$\pm$0.030  & 0.209$\pm$0.015    \\
                 &         & 60002022014 & 2013-09-04 & 12282 & 0.151$\pm$0.019  & 0.074$\pm$0.043  & 0.079$\pm$0.024    \\
\hline
\end{tabular}
}
\end{table*}

\begin{table*}
\caption{\label{ffsrq}Characteristics of flux variations of FSRQs}
\centering
\scalebox{1.1}{
\begin{tabular}{@{}llllllll@{}}
\hline
Name  & Type & OBS ID  & Obs.date  &  Exposure & \multicolumn{3}{c}{$F_{\rm{var}} \pm err(F_{\rm{var}})$}  \\
      &     &        &          &  (secs)      &  3$-$10 keV     & 10$-$79 keV     &  3$-$79 keV     \\
\hline
3C 273         & FSRQ & 00015013001 & 2012-07-02 & 2573   & 0.341$\pm$0.013 & 0.344$\pm$0.019 & 0.341$\pm$0.012  \\
               &      & 00015016001 & 2012-07-02 & 2990   & 0.140$\pm$0.018 & 0.181$\pm$0.028 & 0.164$\pm$0.017  \\
               &      & 10002020001 & 2012-07-14 & 244003 & 0.068$\pm$0.002 & 0.075$\pm$0.002 & 0.072$\pm$0.001  \\
               &      & 10012007001 & 2012-07-13 & 4530   & 0.090$\pm$0.013 & 0.099$\pm$0.021 & 0.096$\pm$0.013  \\
3C 279         & FSRQ & 60002020002 & 2013-12-16 & 39594  & 0.104$\pm$0.009 & 0.102$\pm$0.014 & 0.098$\pm$0.009  \\
               &      & 60002020004 & 2013-12-31 & 42810  & 0.164$\pm$0.007 & 0.174$\pm$0.010 & 0.173$\pm$0.006  \\
PKS 2149$-$306 & FSRQ & 60001099004 & 2014-04-18 & 44167  & 0.092$\pm$0.008 & 0.055$\pm$0.011 & 0.120$\pm$0.007  \\
\hline
\end{tabular}
}
\end{table*}


To check for the robustness of the differences 
in the weighted mean $F_{\rm{var}}$ values of different classes of AGN, 
we carried out two non-parametric statistical tests, namely
the Mann-Whitney U test (hereafter referred to as the U test) and the 
Kolmogorov-Smirnov test (hereafter referred to as the KS test).
The U-test is based on the rank of observations rather than the 
observations themselves. It allows two groups or conditions to be 
compared without making the assumption that the values are 
normally distributed. The null hypothesis that is tested in U-test
is that the distribution of  $F_{\rm{var}}$ values of any two classes of AGN 
(or between different energy bands in a particular class of AGN) that is 
compared is identical. The null hypothesis is rejected (at a particular
level of confidence) if the U-statistics is less than the critical
U-value ($U_{\mathrm{crit}}$). The KS test similar to the U-test is also a non-parametric test that 
can determine if two data sets differ significantly. In this statistics
the cumulative distribution of the two data sets that is to be tested
are plotted and KS test uses the maximum vertical deviation
between the two curves to give the statistics D. The null hypothesis
here is that the two data sets that are compared are from the same
distribution. This null hypothesis is rejected if D is greater than
the critical D value ($D_{\mathrm{crit}}$). The critical values $U_{\mathrm{crit}}$ and
$D_{\mathrm{crit}}$ are evaluated at the 5\% confidence level. In Table \ref{fbl_stat} we give the results
of the two test statistics along with their corresponding P values
for the various comparisons that are studied for the different
classes of AGN. 
From Table \ref{fbl_stat} it is evident that both U and KS tests
reject the null hypothesis of any differences in the variability
properties between FSRQs-BL Lacs and Sy1-Sy2 galaxies in 
soft, hard and total energy bands. However, comparing Seyfert galaxies 
(including Seyfert 1 and 2) and blazars (that include FSRQs and BL Lacs), 
statistical tests clearly indicate that the blazar sources in our sample are 
more variable than Seyfert 
galaxies, in all the three energy bands investigated here, namely soft, 
hard and the full NuSTAR band.
\subsection{Flux variations between soft and hard bands}
To compare the relation between flux variations in the soft (3$-$10 keV)  and hard (10$-$79 keV) X-ray bands, 
$F_{\rm{var}}$ was evaluated for the variable sources in the soft and hard bands. In general, for the different types of
AGN studied in this work, the derived values of $F_{\rm{var}}$ do not show an one 
to one correspondence between the two energy bands as evident from 
Figure~\ref{correlation}. In the case of BL Lac objects, we have a total of 30 sets of observations
on 3 objects. Considering these observations in total, we find that for 
90\% of the observations, the amplitude of variations
in the hard band is larger than the variations in the soft band.  The 
weighted mean
$F_{\rm{var}}$ for BL Lac objects in the hard and soft bands are 0.338 $\pm$ 0.140
and 0.342 $\pm$ 0.177 respectively. Statistical analysis 
by both U and KS tests provides 
no evidence to suggest the variability pattern between 
soft and hard bands in BL Lacs are different. 
FSRQs too show similar amplitude of variations within 
the errors between hard and soft bands and
have weighted mean 
$F_{\rm{var}}$ of 0.082 $\pm$ 0.033 and 0.083 $\pm$ 0.044 in hard and
soft bands respectively. 
Among the 
28 sets of observations on 17 Seyfert 1 galaxies, in about 36\% of
observations the amplitude of flux variations in the hard band are larger than those in 
the soft band. Similarly, in Seyfert 2 galaxies in 13/36 sets of observations, 
the hard band shows a higher variability amplitude compared to the soft band. 
However, based on statistical analysis (Table \ref{fbl_stat}) we find no 
difference
in the variability characteristics between soft and hard bands in Seyfert 1 
and 2 galaxies.
For NLSy1 galaxies, we have 14 sets of observations on 7 objects. 
Considering all the 7 objects together, both U and KS tests show that there is 
no significant difference in the  $F_{\rm{var}}$ values between soft and 
hard bands. Of these, we find that in 3 sets
of observations ($\sim$ 20\%) the variability amplitude 
is larger in the hard band than in the soft
band pertaining to the sources 1H0323$+$342, MCG $+$04$-$22$-$042 and PDS 456.  

\begin{figure*}
\includegraphics[width=15cm,height=13cm]{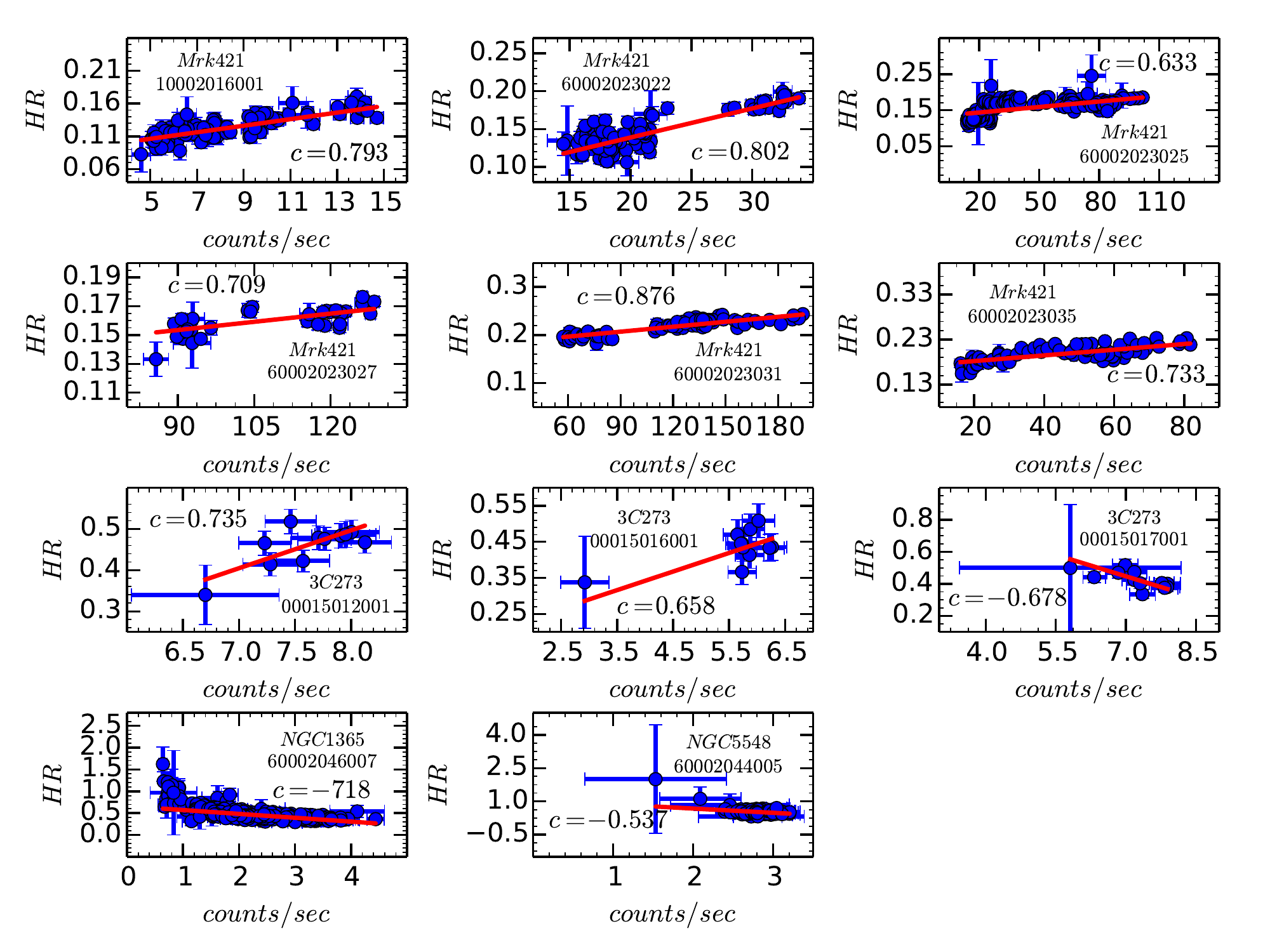}
\caption{\label{hardness}Hardness ratio plotted as a function of 
count rate in the 3$-$79 keV energy range. Red solid
line is the weighted linear least squares fit to the data. The name of the
source, the OBSID and the correlation coefficient are given in each panel.}
\end{figure*}

\begin{table*}
\caption{\label{fsy1}Characteristics of flux variations of Seyfert 1 galaxies}
\centering
\scalebox{1.1}{
\begin{tabular}{@{}llllllll@{}}
\hline
Name  & Type & OBS ID  & Obs.date  &  Exposure & \multicolumn{3}{c}{$F_{\rm{var}} \pm err(F_{\rm{var}})$}  \\
      &     &        &          &  (secs)      &  3$-$10 keV     & 10$-$79 keV     &  3$-$79 keV     \\
\hline
Mrk 335              & Sy1 & 60001041002 & 2013-06-13 & 21299  & 0.190$\pm$0.026  & 0.292$\pm$0.032 & 0.204$\pm$0.022  \\
                     &     & 60001041003 & 2013-06-13 & 21525  & 0.179$\pm$0.025  & 0.045$\pm$0.034 & 0.188$\pm$0.023  \\
                     &     & 60001041005 & 2013-06-25 & 93028  & 0.157$\pm$0.009  & 0.045$\pm$0.013 & 0.114$\pm$0.008  \\
3C 120               & Sy1 & 60001042002 & 2013-02-06 & 21606  & 0.047$\pm$0.006  & 0.035$\pm$0.010 & 0.036$\pm$0.006  \\
                     &     & 60001042003 & 2013-02-06 & 127731 & 0.074$\pm$0.003  & 0.017$\pm$0.005 & 0.044$\pm$0.003  \\
Mrk 9                & Sy1 & 60061326002 & 2013-10-29 & 23310  & 0.066$\pm$0.026  & 0.104$\pm$0.035 & 0.113$\pm$0.024  \\
MCG $-$05$-$23$-$16  & Sy1 & 10002019001 & 2012-07-16 & 33927  & 0.088$\pm$0.003  & 0.074$\pm$0.005 & 0.082$\pm$0.003  \\
                     &     & 60001046002 & 2013-06-03 & 160478 & 0.104$\pm$0.003  & 0.097$\pm$0.005 & 0.096$\pm$0.003  \\
3C 227               & Sy1 & 60061329004 & 2014-02-26 & 12064  & 0.036$\pm$0.018  & 0.158$\pm$0.023 & 0.066$\pm$0.016  \\
NGC 3516             & Sy1 & 60002042004 & 2014-07-11 & 72089  & 0.053$\pm$0.010  & 0.122$\pm$0.014 & 0.060$\pm$0.009  \\   
Mrk 732              & Sy1 & 60061208002 & 2013-06-11 & 26359  & 0.043$\pm$0.013  & 0.041$\pm$0.019 & 0.085$\pm$0.013  \\
NGC 4151             & Sy1 & 60001111002 & 2012-11-12 & 21864  & 0.049$\pm$0.003  & 0.058$\pm$0.004 & 0.050$\pm$0.002  \\
                     &     & 60001111003 & 2012-11-12 & 57036  & 0.099$\pm$0.004  & 0.081$\pm$0.003 & 0.069$\pm$0.003  \\
                     &     & 60001111005 & 2012-11-14 & 61531  & 0.099$\pm$0.002  & 0.082$\pm$0.003 & 0.088$\pm$0.002  \\ 
Mrk 231              & Sy1 & 60002025004 & 2013-05-09 & 28557  & 0.215$\pm$0.044  & 0.149$\pm$0.062 & 0.047$\pm$0.041  \\
MCG $-$06$-$30$-$15  & Sy1 & 60001047002 & 2013-01-29 & 23270  & 0.225$\pm$0.007  & 0.181$\pm$0.012 & 0.199$\pm$0.007  \\
                     &     & 60001047003 & 2013-01-30 & 127232 & 0.314$\pm$0.003  & 0.263$\pm$0.005 & 0.289$\pm$0.003  \\  
                     &     & 60001047005 & 2013-02-02 & 29646  & 0.173$\pm$0.007  & 0.140$\pm$0.010 & 0.154$\pm$0.006  \\  
NGC 5506             & Sy1 & 60061323002 & 2014-04-01 & 56585  & 0.072$\pm$0.004  & 0.052$\pm$0.006 & 0.063$\pm$0.004  \\ 
NGC 5548             & Sy1 & 60002044006 & 2013-09-10 & 51460  & 0.066$\pm$0.005 & 0.094$\pm$0.007 & 0.066$\pm$0.005  \\  
                     &     & 60002044008 & 2013-12-20 & 50103  & 0.057$\pm$0.006  & 0.054$\pm$0.008 & 0.073$\pm$0.006  \\ 
Mrk 290              & Sy1 & 60061266002 & 2013-11-14 & 25012  & 0.087$\pm$0.013  & 0.032$\pm$0.020 & 0.023$\pm$0.013  \\  
                     &     & 60061266004 & 2013-11-27 & 26348  & 0.038$\pm$0.014  & 0.173$\pm$0.021 & 0.057$\pm$0.013  \\ 
3C 390.3             & Sy1 & 60001082002 & 2013-05-24 & 23643  & 0.042$\pm$0.006  & 0.062$\pm$0.010 & 0.053$\pm$0.006  \\
                     &     & 60001082003 & 2013-05-24 & 47559  & 0.020$\pm$0.005  & 0.072$\pm$0.007 & 0.020$\pm$0.005  \\
IRAS F21318$-$2739   & Sy1 & 60061306002 & 2013-10-22 & 19809  & 0.045$\pm$0.017  & 0.139$\pm$0.029 & 0.026$\pm$0.017  \\
NGC 7582             & Sy1 & 60061318002 & 2012-08-31 & 16463  & 0.152$\pm$0.018  & 0.108$\pm$0.016 & 0.137$\pm$0.013  \\
IC 4329A              & Sy1 & 60001045002 & 2012-08-12 & 162399 & 0.118$\pm$0.002  & 0.103$\pm$0.004 & 0.109$\pm$0.002  \\
\hline
\end{tabular}
}
\end{table*}

\begin{table*}
\caption{\label{fsy2}Characteristics of flux variations of Seyfert 2 galaxies}
\centering
\begin{tabular}{@{}llllllll@{}}
\hline
Name  & Type & OBS ID  & Obs.date  &  Exposure & \multicolumn{3}{c}{$F_{\rm{var}} \pm err(F_{\rm{var}})$}  \\
      &     &        &          &  (secs)      &  3$-$10 keV     & 10$-$79 keV     &  3$-$79 keV     \\
\hline
NGC 513               & Sy2 & 60061012002 & 2013-02-16 & 16040 & 0.113$\pm$0.027 & 0.097$\pm$0.033 & 0.115$\pm$0.023  \\
NGC 788               & Sy2 & 60061018002 & 2013-01-28 & 15411 & 0.088$\pm$0.031 & 0.110$\pm$0.024 & 0.093$\pm$0.020  \\ 
NGC 1068              & Sy2 & 60002030002 & 2012-12-18 & 57851 & 0.057$\pm$0.010 & 0.054$\pm$0.013 & 0.027$\pm$0.008  \\ 
                      &     & 60002030004 & 2012-12-20 & 48560 & 0.035$\pm$0.010 & 0.071$\pm$0.013 & 0.065$\pm$0.009  \\ 
                      &     & 60002030006 & 2012-12-21 & 19461 & 0.052$\pm$0.016 & 0.120$\pm$0.020 & 0.087$\pm$0.013  \\ 
NGC 1365              & Sy2 & 60002046002 & 2012-07-25 & 36258 & 0.276$\pm$0.010 & 0.311$\pm$0.034 & 0.215$\pm$0.019  \\ 
                      &     & 60002046003 & 2012-07-26 & 40588 & 0.157$\pm$0.011 & 0.103$\pm$0.011 & 0.130$\pm$0.008  \\ 
                      &     & 60002046005 & 2012-12-24 & 66297 & 0.202$\pm$0.005 & 0.161$\pm$0.007 & 0.172$\pm$0.004  \\
                      &     & 60002046007 & 2013-01-23 & 73650 & 0.431$\pm$0.004 & 0.255$\pm$0.007 & 0.338$\pm$0.004  \\  
                      &     & 60002046009 & 2013-02-12 & 69877 & 0.397$\pm$0.006 & 0.218$\pm$0.009 & 0.280$\pm$0.006  \\
MCG $+$03$-$13$-$01   & Sy2 & 60061051002 & 2014-03-18 & 20088 & 0.139$\pm$0.045 & 0.047$\pm$0.040 & 0.122$\pm$0.031  \\
XSS J05054$-$2348     & Sy2 & 60061056002 & 2013-08-21 & 21161 & 0.035$\pm$0.010 & 0.077$\pm$0.013 & 0.060$\pm$0.009  \\
IRAS 05189$-$2524     & Sy2 & 60002027002 & 2013-02-20 & 23141 & 0.060$\pm$0.023 & 0.236$\pm$0.038 & 0.177$\pm$0.023   \\
                      &     & 60002027004 & 2013-10-02 & 25370 & 0.213$\pm$0.026 & 0.181$\pm$0.045 & 0.227$\pm$0.025  \\
NGC 2110              & Sy2 & 60061061004 & 2013-02-14 & 12019 & 0.028$\pm$0.008 & 0.028$\pm$0.011 & 0.073$\pm$0.007  \\
IRAS 07378$-$3136     & Sy2 & 60061351002 & 2014-04-20 & 23952 & 0.077$\pm$0.020 & 0.075$\pm$0.020 & 0.038$\pm$0.015  \\
Mrk 1210              & Sy2 & 60061078002 & 2012-10-05 & 15447 & 0.115$\pm$0.014 & 0.120$\pm$0.015 & 0.121$\pm$0.011  \\
MCG $+$01$-$24$-$012  & Sy2 & 60061091002 & 2013-04-03 & 12376 & 0.063$\pm$0.014 & 0.076$\pm$0.020 & 0.048$\pm$0.013  \\
                      &     & 60061091004 & 2013-04-10 &  9386 & 0.141$\pm$0.016 & 0.044$\pm$0.022 & 0.111$\pm$0.014  \\
                      &     & 60061091006 & 2013-04-18 & 12178 & 0.171$\pm$0.015 & 0.026$\pm$0.028 & 0.053$\pm$0.016  \\
                      &     & 60061091010 & 2013-05-12 & 15334 & 0.050$\pm$0.011 & 0.050$\pm$0.016 & 0.024$\pm$0.010  \\
                      &     & 60061091012 & 2013-05-22 & 12289 & 0.063$\pm$0.012 & 0.020$\pm$0.017 & 0.038$\pm$0.011  \\
NGC 3079              & Sy2 & 60061097002 & 2013-11-12 & 21542 & 0.126$\pm$0.054 & 0.131$\pm$0.029 & 0.125$\pm$0.026  \\
NGC 4395              & Sy2 & 60061322002 & 2013-05-10 & 19249 & 0.365$\pm$0.015 & 0.281$\pm$0.018 & 0.320$\pm$0.013  \\
Mrk 248               & Sy2 & 60061241002 & 2013-04-21 & 12901 & 0.033$\pm$0.019 & 0.017$\pm$0.025 & 0.036$\pm$0.017  \\
                      &     & 60061241004 & 2013-11-17 & 28909 & 0.040$\pm$0.013 & 0.016$\pm$0.018 & 0.065$\pm$0.011  \\
                      &     & 60061241006 & 2013-11-23 & 23056 & 0.127$\pm$0.015 & 0.115$\pm$0.022 & 0.107$\pm$0.014   \\
NGC 5273              & Sy2 & 60061350002 & 2014-07-14 & 21119 & 0.167$\pm$0.009 & 0.130$\pm$0.013 & 0.153$\pm$0.009  \\
NGC 5674              & Sy2 & 60061337002 & 2014-07-10 & 20671 & 0.073$\pm$0.014 & 0.107$\pm$0.021 & 0.051$\pm$0.013  \\
Mrk 477               & Sy2 & 60061255002 & 2014-05-15 & 18076 & 0.085$\pm$0.022 & 0.027$\pm$0.025 & 0.027$\pm$0.018  \\
NGC 6300              & Sy2 & 60061277002 & 2013-02-25 & 17706 & 0.280$\pm$0.010 & 0.184$\pm$0.012 & 0.226$\pm$0.008  \\
LEDA 3097193          & Sy2 & 60061354002 & 2014-05-19 & 15645 & 0.114$\pm$0.015 & 0.132$\pm$0.019 & 0.133$\pm$0.013  \\
H 1834$-$653          & Sy2 & 60061288002 & 2013-02-24 & 27391 & 0.054$\pm$0.007 & 0.021$\pm$0.009 & 0.057$\pm$0.006  \\
IGR J19473$+$4452  & Sy2 & 60061292002 & 2012-11-06 & 18214 & 0.024$\pm$0.017 & 0.069$\pm$0.024 & 0.057$\pm$0.015  \\
MCG $+$07$-$41$-$003  & Sy2 & 60001083004 & 2013-03-01 & 20715 & 0.019$\pm$0.006 & 0.040$\pm$0.009 & 0.029$\pm$0.006  \\
IGR J20187$+$4041     & Sy2 & 60061297002 & 2013-12-21 & 20967 & 0.099$\pm$0.017 & 0.065$\pm$0.021 & 0.080$\pm$0.014  \\
\hline
\end{tabular}
\end{table*}

\begin{figure*}
\centering
\hbox{
      \includegraphics[width=7cm,height=6cm]{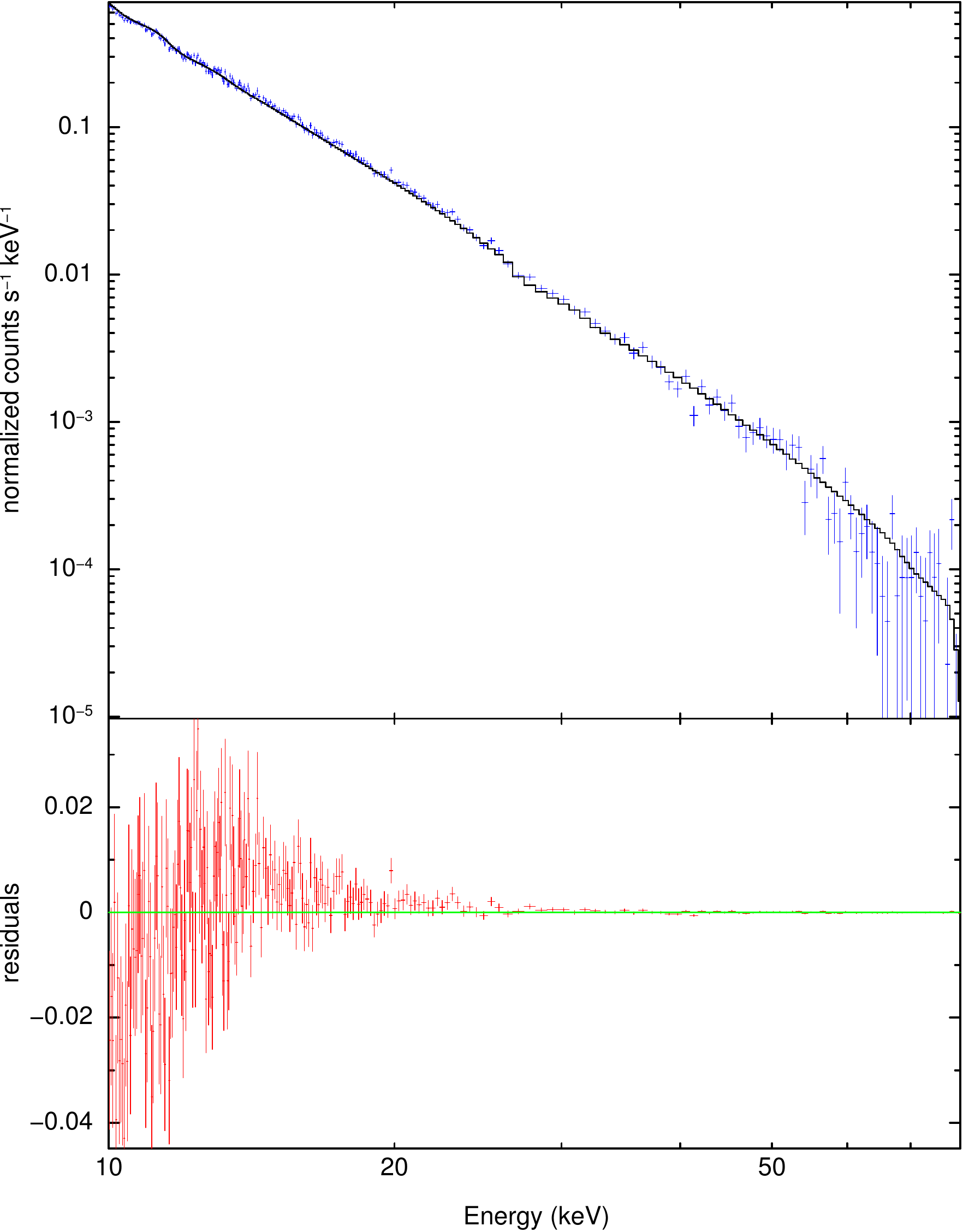}
      \hspace*{2.0cm}\includegraphics[width=7cm,height=6cm]{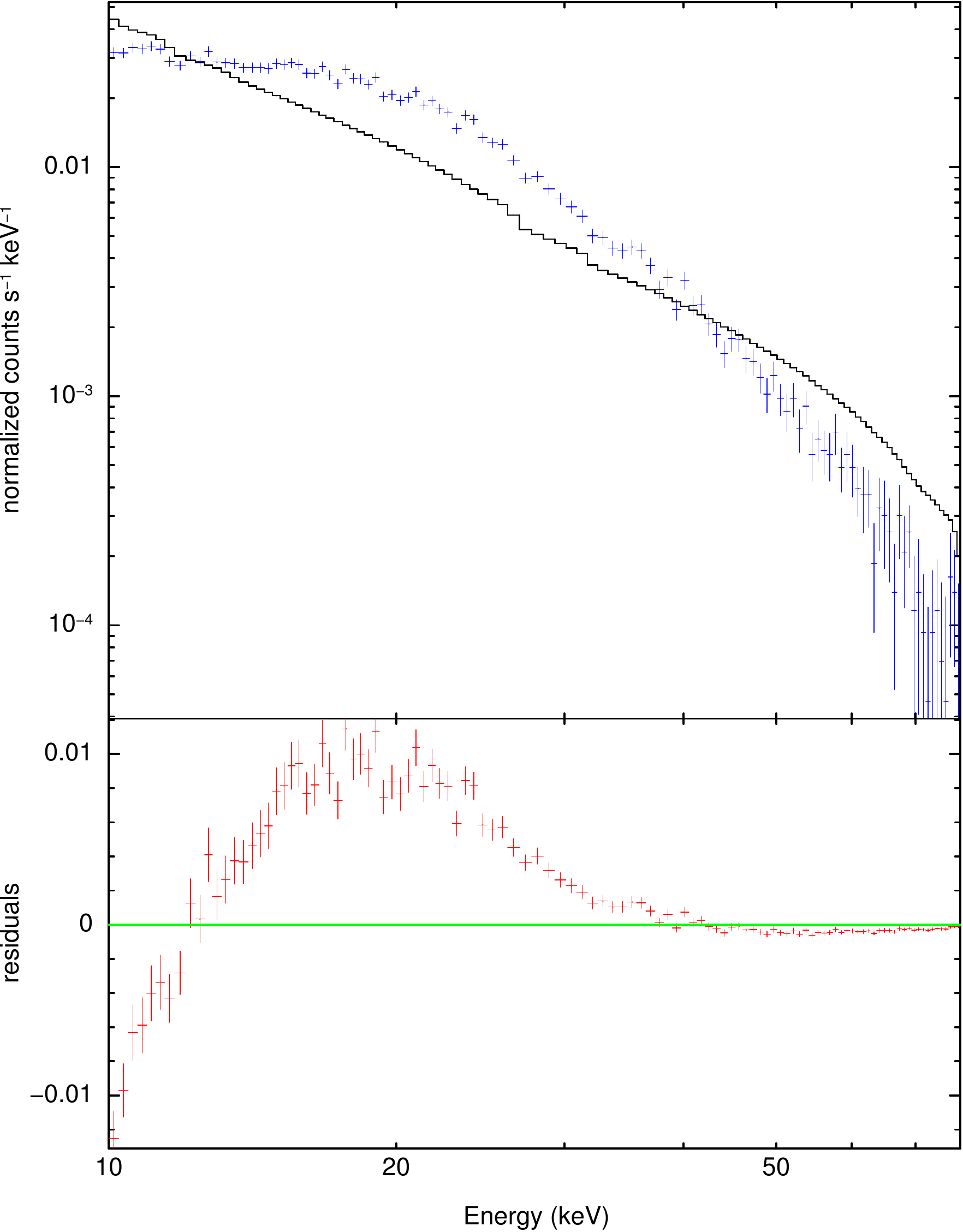}
     }
\caption{\label{spectra} 
Hard X-ray spectrum of the BL Lac Mrk 421 (left) and the Seyfert 1 galaxy
PKS 1409$-$651 (right) fitted
with a power law model along with the residuals.}
\end{figure*}

\subsection{Flux variability time scale}
Knowledge of the time scales on which the X-ray flux varies is very 
important
as it can put constraints on the size of the emission region.
For sources that have shown flux variations,  we scan their light curves 
in the energy range 3$-$10 keV and 10$-$79 keV to find the time scale of flux 
variations. 
For this we calculated the flux doubling time/halving time  defined as
\begin{equation}
\centering
F(t)=F(t_{\rm{0}})\times2^{(t-t_{\rm{0}})/\tau}
\end{equation}
here, $\tau$ is the characteristic flux doubling/halving time scale  
and $F(t_{\rm{0}})$ and $F(t)$ are 
values of the fluxes at time $t_{\rm{0}}$ and $t$ respectively. This time 
scale is evaluated by imposing the condition
that the difference between the fluxes at times  $t_{\rm{0}}$ and $t$ is 
greater than $3\sigma$ \citep{2011A&A...530A..77F}. The best fit
values obtained on fitting Equation 5 to the data are given in
Table~\ref{timescales}.
The quoted uncertainties in $\tau$ are the 1 $\sigma$ errors. 
A total of 11 sources
are found to have flux doubling/halving time scale less than 10 min. 
Such behaviour is not restricted to one AGN type 
and is seen in all types of AGN. Among the Seyfert 2 galaxies
NGC 1365 has shown flux doubling/halving time less than 10 min in all
three epochs it has been observed. 

\begin{figure}
\includegraphics[width=9cm,height=7cm]{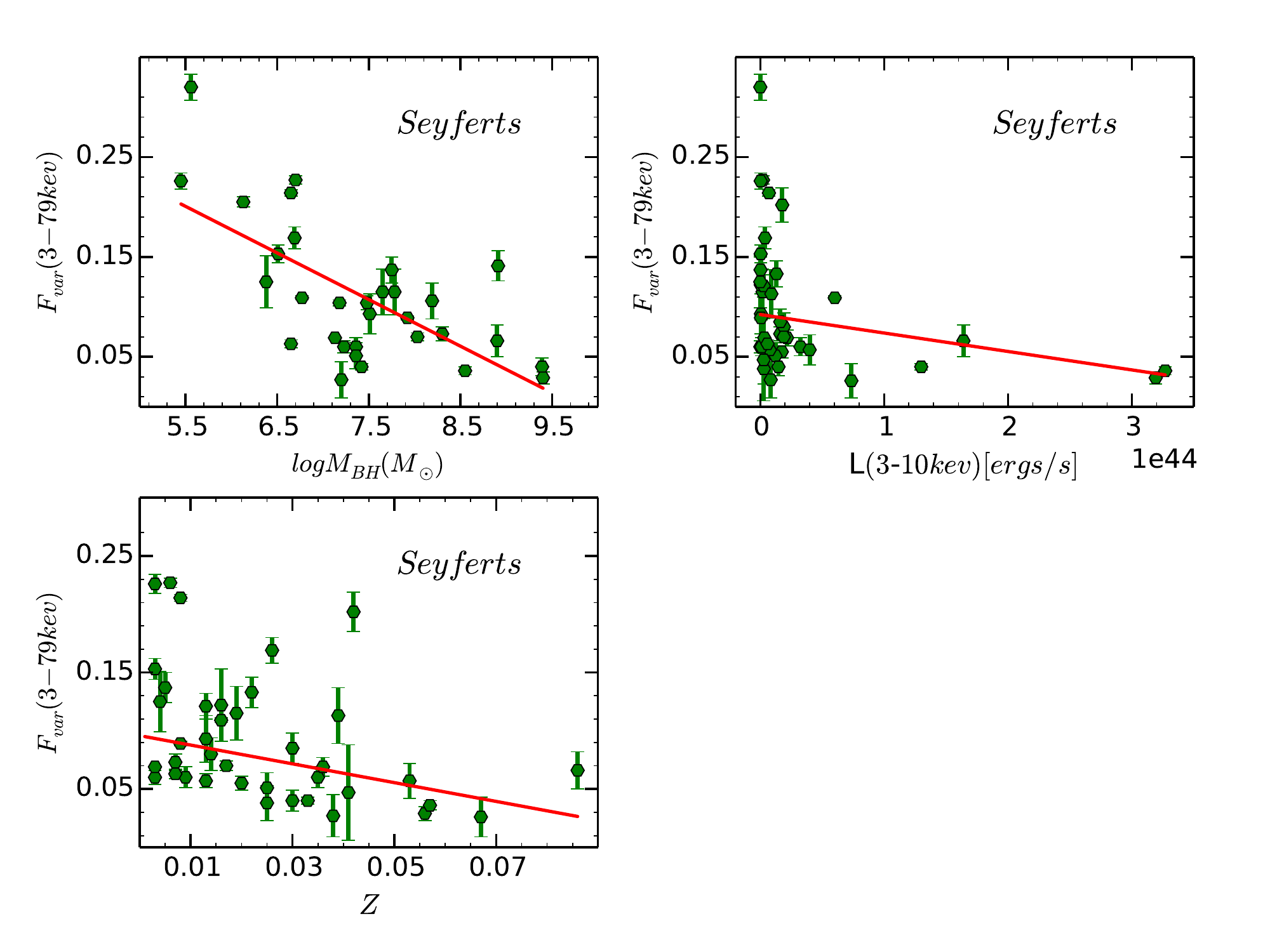}
\caption{\label{Fvar_correlation}Correlation of $F_{\rm{var}}$ with BH mass, redshift and luminosity in the 3$-$10 keV band.}
\end{figure}

\begin{table*}
\caption{\label{fnlsy1}Characteristics of flux variations of NLSy1 galaxies}
\centering
\begin{tabular}{@{}llllllll@{}}
\hline
Name  & Type & OBS ID  & Obs.date  &  Exposure & \multicolumn{3}{c}{$F_{\rm{var}} \pm err(F_{\rm{var}})$}  \\
      &     &        &          &  (secs)      &  3$-$10 keV     & 10$-$79 keV     &  3$-$79 keV     \\
\hline
1H 0323$+$342        & NLSy1 & 60061360002 & 2014-03-15 & 101633 & 0.141$\pm$0.010 & 0.172$\pm$0.017 & 0.138$\pm$0.010  \\
MCG $+$04$-$22$-$042  & NLSy1 & 60061092002 & 2012-12-26 &  18845 & 0.020$\pm$0.009 & 0.085$\pm$0.015 & 0.081$\pm$0.009  \\
NGC 4051             & NLSy1 & 60001050002 & 2013-06-17 &   9434 & 0.225$\pm$0.014 & 0.058$\pm$0.024 & 0.168$\pm$0.014  \\
                     &       & 60001050003 & 2013-06-17 &  45737 & 0.341$\pm$0.006 & 0.214$\pm$0.010 & 0.277$\pm$0.006  \\
                     &       & 60001050005 & 2013-10-09 &  10202 & 0.164$\pm$0.016 & 0.064$\pm$0.024 & 0.128$\pm$0.015  \\
                     &       & 60001050006 & 2013-10-09 &  49621 & 0.352$\pm$0.009 & 0.198$\pm$0.012 & 0.273$\pm$0.008  \\
                     &       & 60001050008 & 2014-02-16 &  56683 & 0.226$\pm$0.005 & 0.132$\pm$0.009 & 0.180$\pm$0.005  \\
IGR J14552$-$5133    & NLSy1 & 60061259002 & 2013-09-19 &  21943 & 0.170$\pm$0.017 & 0.148$\pm$0.027 & 0.160$\pm$0.017  \\
MCG $+$05$-$40$-$026 & NLSy1 & 60061276002 & 2013-12-19 &  20999 & 0.147$\pm$0.019 & 0.111$\pm$0.029 & 0.073$\pm$0.018  \\
PDS 456              & NLSy1 & 60002032010 & 2014-02-26 & 109717 & 0.144$\pm$0.020 & 0.161$\pm$0.037 & 0.180$\pm$0.022  \\
IGR J21277$+$5656    & NLSy1 & 60001110002 & 2012-11-04 &  49202 & 0.143$\pm$0.006 & 0.122$\pm$0.010 & 0.138$\pm$0.006  \\
                     &       & 60001110003 & 2012-11-05 &  28765 & 0.096$\pm$0.007 & 0.019$\pm$0.012 & 0.054$\pm$0.007  \\
                     &       & 60001110005 & 2012-11-06 &  74583 & 0.139$\pm$0.004 & 0.112$\pm$0.007 & 0.124$\pm$0.004  \\
                     &       & 60001110007 & 2012-11-08 &  42110 & 0.128$\pm$0.006 & 0.085$\pm$0.009 & 0.101$\pm$0.006  \\
\hline
\end{tabular}
\end{table*}

\subsection{Time delay between flux variations in hard and soft bands}
To quantify the degree of correlation between flux variations in soft and hard
bands we used a model based approach using the code JAVELIN
\footnote{http://www.astronomy.ohio-state.edu/~yingzu/codes.html}, a Python
based implementation of SPEAR (Stochastic Process Estimation for 
AGN Reverberation; \citealt{2011ApJ...735...80Z,2013ApJ...765..106Z}).
JAVELIN has been designed to improve lag measurements between continuum and line variations in AGN by modelling the light curves using a damped random walk 
process (DRWP; \citealt{2009ApJ...698..895K,2010ApJ...721.1014M}). 
It then
compares the simulated light curves with the observed ones to find the lag
and this has been found to work well 
\citep{2014MNRAS.445.3055P}. We  applied the JAVELIN method to all the sources
in our sample. For all the sources 
the variations between soft and hard bands are consistent with zero lag. 
Several studies do exist in the literature on time lags between flux variations
in different X-ray bands. For blazars in general using data in the 
0.3$-$ 10 keV band, soft lags (lower energy variations lagging the higher
energy variations), hard lags (higher energy variations lagging the
lower energy variation) and zero lags have been observed 
\citep{2003A&A...402..929B,2004ApJ...601..165F,2004A&A...424..841R,2006ApJ...637..699Z,
2017ApJ...834....2A}. 
This indicates
the complex spectral and temporal characteristics of blazars that show
different behaviour at different times. Observations of few Seyfert galaxies
in the hard X-ray band by SWIFT/BAT in 2 days binned light curve
did not show any delay between 20$-$50 and 50$-$100 keV bands
\citep{2012A&A...537A..87C}.

\begin{table}
\caption{\label{afvar}Weighted mean  variability characteristics of different classes of AGN. N1 and
N2 represent the number of objects and the number of observations respectively.}
\begin{tabular}{@{}llllll@{}}
\hline
Type     & N1     & N2  & \multicolumn{3}{c} {avg$(F_{\rm{var}} \pm err(F_{\rm{var}}))$}  \\
         &  &       & 3$-$10 keV     & 10$-$79 keV     &  3$-$79 keV     \\
\hline
BL\_Lac  &   3      &     30        & 0.342$\pm$0.177  & 0.338$\pm$0.140 & 0.310$\pm$0.138   \\
FSRQ     &   3      &     7         & 0.083$\pm$0.044  & 0.082$\pm$0.033 & 0.078$\pm$0.028   \\
Sy1      &  17      &     28        & 0.108$\pm$0.068  & 0.089$\pm$0.052 & 0.092$\pm$0.061   \\
Sy2      &  23      &     36        & 0.191$\pm$0.156  & 0.120$\pm$0.082 & 0.149$\pm$0.108   \\
NLSy1    &   7      &     14        & 0.177$\pm$0.085  & 0.122$\pm$0.052 & 0.151$\pm$0.065   \\
LSP      &   3      &     7         & 0.083$\pm$0.044  & 0.082$\pm$0.033 & 0.078$\pm$0.028   \\
HSP      &   3      &     30        & 0.342$\pm$0.177  & 0.338$\pm$0.140 & 0.310$\pm$0.138    \\
Blazars  &   6      &     37        & 0.321$\pm$0.184  & 0.258$\pm$0.167 & 0.241$\pm$0.158  \\
Seyferts &  40      &     64        & 0.124$\pm$0.098  & 0.095$\pm$0.061 & 0.104$\pm$0.076   \\
\hline
\end{tabular}
\end{table}
 
\subsection{Duty cycle of flux variations}
To characterize the incidence of observability of X-ray variations
in different classes of AGN, we have calculated the duty cycle of 
X-ray flux variations using the definition of \cite{1999A&AS..135..477R}. 
Since only about 65\% of the sources in our sample has shown variations
exceeding the measurement noise characterised by $F_{\rm{var}}$, 
duty cycles are estimated not as a fraction of the variable objects 
within a given class, but as the ratio of the time over which the objects 
of a given class are found to vary to the total time of observations
carried out on each objects in the class. An approach of this kind
will take into account the fact that all AGN do not show variations at 
all times.
Also, as the duration of observation is not the same for all the objects, 
the contribution to the duty cycle has been weighted by the number
of times, as well as the duration each source was observed. Duty cycle is defined as 

\begin{equation}
\centering
DC={\sum_{i=1}^{n}N_{i}(1/\Delta t_{i})\over\sum_{i=1}^{n}(1/\Delta t_{i})} \times 100  \%
\end{equation}

Here, $\Delta t_{i}=\Delta t_{0}(1+z)^{-1}$, is the duration corrected for 
cosmological redshift of 
the source observed
 and  $N_{i}$ equals 1 if the object is variable during the period of 
observations $\Delta t_{i}$ and 0 otherwise.  We find NLSy1 galaxies show the
highest DC of 87.14\% followed by BL Lacs (82.44 \%), Seyfert galaxies 
(Seyfert 1 and 2 galaxies show similar DC of variability of 
55.87 \% and 56.47\% respectively) and FSRQs (22.79\%). Most
of the BL Lacs belong to the HSP category of AGN and FSRQs are in general
LSP sources. In the X-ray band, HSPs are known to be more variable
in the X-ray band than LSPs, because, in the former the X-ray spectrum
falls in the synchrotron region of the SED, while in the latter, it
falls in the IC region.  Therefore, the difference in the DC of variability
between BL Las and FSRQs is due to the differences in the physical 
processes that contribute to the X-ray emission in the NuSTAR band
in these two classes of AGN.

\begin{table*}
\caption{\label{fbl_stat} Results of statistical tests to compare
the $F_{\rm{var}}$ properties of different classes of AGN}
\centering
\begin{tabular}{@{}lrrrrrrrl@{}} \cline{2-9}
\hline
Parameters  & \multicolumn{4}{c} {Mann-Whitney U test}   & \multicolumn{4}{c} {Kolmogorov-Smirnov test} \\ 
                          &  $U_{\mathrm{obs}}$   & $U_{\mathrm{crit}}$   & Null hypothesis   &   P      &  D       &  $D_{\mathrm{crit}}$ & Null hypothesis     & P \\ \hline
Sy1 \& Sy2 (soft band)    &  471.5   & 358.7  & No                       & 0.66     &  0.1508  & 0.3427  & No                       & 0.836 \\ 
Sy1 \& Sy2 (hard band)    &  492.5   & 358.7  & No                       & 0.88     &  0.1587  & 0.3427  & No                       & 0.787 \\
Sy1 \& Sy2 (full band)    &  446.0   & 358.7  & No                       & 0.43     &  0.2024  & 0.3427  & No                       & 0.491 \\
BL \& FSRQ (soft band)    &  76.5      & 54     & No               & 0.28     &          &         &                          &       \\
BL \& FSRQ (hard band)    &  65      & 54     & No                 & 0.13     &          &         &                          &       \\
BL \& FSRQ (full band)    &  78      & 54     & No                & 0.30     &          &         &                          &       \\
Sy \& blazars (soft band) &  708     &  905   & Yes               & 0.0008 & 0.3792   & 0.2809 & Yes                     & 0.002 \\
Sy \& blazars (hard band) & 521.5    & 905    & Yes              & 0.0001 & 0.4569   & 0.2809 & Yes                     & 0.000 \\
Sy \& blazars (full band) & 553.5    & 905    & Yes              & 0.0001 & 0.4430   & 0.2809 & Yes                     & 0.000 \\
NLSy1 (hard v/s soft)$^{*}$     & 59       & 55     & No      & 0.0767   & 0.4286   & 0.5140 & No   & 0.111 \\
NLSy1 (hard v/s soft)\textdagger     & 33       & 37.6     & Yes   & 0.0264   & 0.5000   & 0.5552 & No       		    & 0.026 \\
FSRQ  (hard v/s soft)     & 23       & 8      & No                     & 0.8966   &          &          &                         &       \\
BL    (hard v/s soft)     & 339      & 316    & No                     & 0.1031   & 0.2667   & 0.3511 & No                      & 0.200 \\
Sy1   (hard v/s soft)     & 387      & 271.8  & No                       & 0.9442   & 0.1071   & 0.3635 & No                      & 0.995 \\
Sy2   (hard v/s soft)     &          & 473.5  &                          & 0.5157   & 0.1111   & 0.3206 & No                      & 0.971 \\ 
NLSy1 \& BLSy1 (soft)     & 102     & 122.0  & Yes                      & 0.0128   & 0.5714   & 0.4452 & Yes                     & 0.002 \\
NLSy1 \& BLSy1 (hard)     & 145      & 122.0  & No                       & 0.177   & 0.3214   & 0.4452 & No                      & 0.237 \\ 
NLSy1 \& blazars (soft)   & 220.5    & 165.6  & No                       & 0.4237   & 0.2896  & 0.4267 & No                      & 0.306 \\
NLSy1 \& blazars (hard)   & 141    & 165.6  & Yes                      & 0.0131   & 0.4054   & 0.4267 & No                      & 0.051 \\ \hline
\hline
\end{tabular}\\
$^{*}$\footnotesize{Considering all NlSy1 galaxies that include radio-quiet and radio-loud sources.}\\
{\textdagger}\footnotesize{Considering only radio-quiet NlSy1 galaxies, that include NGC 4051, \\IGR J14552$-$5113, MCG $+$05$-$40$-$026 and IGR J21277$+$5656.}
\end{table*}

\subsection{Spectral variations} 
The lack of one to one correspondence between the amplitude of flux variations 
in the
soft and hard bands in some sources indicate that they show spectral
variations. 
To further characterise spectral variations, we construct diagrams
of hardness ratio (HR) plotted as a function of total flux in the 
3$-$79 keV energy range. This is a model independent way to study spectral
variations.  HR is estimated using the following relation
\begin{equation}
HR = \frac{F_{hard}} {F_{soft}} 
\end{equation}
where, $F_{hard}$ and $F_{soft}$ refer to the fluxes in the 10$-$79 keV and 
the 3$-$10 keV respectively. For most of the sources we do not find 
a significant correlation between variations in HR and total flux. However, for
some sources we do find a correlation. For those sources, the plot of HR 
as a function of the flux in the 3$-$79 keV band is shown in 
Figure~\ref{hardness}. To quantify the significance of the correlation, we fit 
the observed points in the HR
v/s flux diagram using  a linear function of the form 
$HR = a \times flux_{3-79keV} + b$. During the fit we take into account the 
errors both in HR and flux following \cite{1992nrca.book.....P}.
The results of the fit are given in Table~\ref{lagresults} and they are shown as solid
lines in Figure~\ref{hardness}. Significant spectral variations are seen in 
the BL Lacs object Mrk 421, the FSRQ 3C 273, the Seyfert 1 
galaxy NGC 5548 and the Seyfert 2 galaxy NGC 1365. 
For the BL Lac object Mrk 421 a significant harder when brighter trend 
is seen in 6 epochs of observations.
However, for the FSRQ 3C 273, in the three epochs where a correlation between
HR and total flux is found, on two epochs a harder when brighter
trend is found, while, in one epoch a softer when brighter trend is noticed.
Blazars in general are found
to show a harder when brighter behaviour. Such 
hardening when brightening behaviour more often seen in the HSP category
\citep{1990ApJ...356..432G,1998ApJ...492L..17P}
among other things could be due
to the shift of their broad band SEDs to higher energies
\citep{2003A&A...402..929B}. The behaviour seen in Mrk 421 here is observed
before as well \citep{1996ApJ...470L..89T}. 
In the radio-quiet category, two Seyfert galaxies, namely NGC 5548 and
NGC 1365 showed spectral variations with their spectra becoming softer
with increasing brightness. This trend is also known in other
Seyfert galaxies based on observations from RXTE \citep{2009MNRAS.399.1597S}, 
Swift/XRT \citep{2016MNRAS.459.3963C} and
Swift/BAT \citep{2012A&A...537A..87C}, which is expected
in various models \citep{1997ApJ...476..620H,2009A&ARv..17...47T}.
We conclude that among the sample of sources studied here,
only some sources do show spectral variations. The hard
X-ray band in Mrk 421 is primarily dominated by the 
non-thermal continuum (not
necessarily true for the Seyfert galaxies NGC 5548 and NGC 1365), 
while in the soft X-ray band there can be contribution from different 
processes such as power law continuum, soft excess, neutral absorption and 
absorption features from warm absorbers. Thus, the harder when brighter 
trend seen in Mrk 421 is most likely due to 
changes in the power law component in the relativistic jets of these sources.

\begin{table*}
\centering
\caption{\label{timescales}The shortest flux doubling/halving time in minutes and its significance.}
\begin{tabular}{@{}lllrrrrl@{}}
\hline
Name		& Type		& OBSID		&  $\tau$		& Sig.   	& $\tau$	 &  Sig.  \\
                &               &               &   (3$-$10 keV)        &               &  (10$-$79 keV) &        \\
                &               &               &     (min.)            &               &       (min.)   &        \\
\hline
3C 120               & Sy1   & 60001042002  &  17.36 $\pm$  5.60 &  3.11 &  5.25 $\pm$  2.01 &  3.33  \\
MCG $+$07$-$41$-$003 & Sy2   & 60001083002  &  16.82 $\pm$  5.49 &  3.07 &  9.13 $\pm$ 2.63  &  3.56  \\ 
NGC 4051          & NLSy1 & 60001050008  &   6.96 $\pm$  1.25 &  5.65 &  5.70 $\pm$  1.48 &  3.91   \\
NGC 4151          & Sy1   & 60001111005  &  16.42 $\pm$  5.78 &  3.08 & 23.53 $\pm$  6.36 &  3.70   \\
Mrk 421           & BLLac & 10002015001  &  23.40 $\pm$  6.66 &  3.52 & 43.40 $\pm$ 13.44 &  3.85   \\
                  &       & 60002023006  &  22.32 $\pm$  6.80 &  3.29 &  3.04 $\pm$  1.23 &  3.75   \\
                  &       & 60002023022  & 113.64 $\pm$ 15.26 &  7.71 & 96.20 $\pm$ 26.51 &  3.78   \\
                  &       & 60002023025  &  37.08 $\pm$  8.69 &  4.27 & 16.59 $\pm$  4.44 &  3.98   \\
                  &       & 60002023027  &  55.84 $\pm$  4.72 & 11.57 & 47.44 $\pm$  8.28 &  5.58   \\
                  &       & 60002023031  &  34.13 $\pm$  3.47 &  9.84 & 34.72 $\pm$  9.89 &  3.43   \\
                  &       & 60002023033  &  55.96 $\pm$ 16.92 &  3.31 & 22.33 $\pm$  6.48 &  3.45   \\
                  &       & 60002023035  &  25.11 $\pm$  2.54 &  9.88 & 25.07 $\pm$  5.60 &  4.48   \\
3C 273             & FSRQ  & 10002020001  &   20.64 $\pm$  6.36 &  3.25 &  11.06 $\pm$  2.45 & 4.52   \\
IC 4329A           & Sy1   & 60001045002  &  22.11 $\pm$  7.27 &  3.05 & 14.12 $\pm$  4.19 &  3.38   \\
MCG $-$05$-$23$-$16       & Sy1   & 60001046002  &  99.58 $\pm$ 23.12 &  4.23 &  3.35 $\pm$  1.45 &  3.89  \\
MCG $-$06$-$30$-$15       & Sy1   & 60001047003  &  10.45 $\pm$  3.18 &  3.30 &  7.89 $\pm$  2.53 &  3.15   \\
PDS 456           & NLSy1 & 60002032002  &   4.94 $\pm$  1.45 &  3.27 &  1.99 $\pm$  0.85 &  3.34   \\
IGR J21277$+$5656 & NLSy1 & 60001110002  &  10.36 $\pm$  2.84 &  3.68 & 32.59 $\pm$  8.63 &  3.56   \\
                  &       & 60001110005  &   5.78 $\pm$  1.93 &  3.79 & 51.47 $\pm$ 17.07 &  3.04   \\
NGC 1068          & Sy2   & 60002030004  &  17.63 $\pm$  8.07 &  3.98 &  5.12 $\pm$  1.72 &  3.02   \\
NGC 1365          & Sy2   & 60002046005  &  10.49 $\pm$  2.45 &  4.32 &  5.36 $\pm$  2.17 &  3.14   \\
                  &       & 60002046007  &   8.45 $\pm$  2.16 &  3.94 &  7.70 $\pm$  2.83 &  3.04   \\
                  &       & 60002046009  &  10.56 $\pm$  3.16 &  3.35 &  7.76 $\pm$  2.39 &  3.28   \\
NGC 1052          & Sy2   & 60061027002  &   3.53 $\pm$  1.45 &  3.10 &  1.86 $\pm$  1.02 &  3.67   \\
NGC 4395          & Sy2   & 60061322002  &   3.08 $\pm$  0.44 &  7.33 &  4.35 $\pm$  1.08 &  4.13   \\

\hline
\end{tabular}
\end{table*}

\section{Discussion}
X-ray flux variations in AGN can be described by a wide variety of physical
processes. In the radio-quiet category of objects hard X-rays can be 
produced by the Comptonization of soft disk photons in the hot plasma
above the accretion disk \citep{1993ApJ...413..507H}, whereas, in 
the radio-loud category of AGN, hard X-rays 
are dominated by inverse Compton scattering of 
synchrotron photons by relativistic electrons in the jet via 
Synchrotron Self Compton process and synchrotron radiation. Given the 
various methods of the production of hard X-rays in AGN, there can be
different causes for X-ray flux variations. 
In this section we discuss the
results of our variability analysis, the potential causes of such variations,
the nature of X-ray variations as well as their relation to various 
physical properties of the sources.
\subsection{Flux variability}
From the results of our variability analysis on a large sample of AGN
belonging to different categories, we find that a major fraction of
65\% of the sources in our sample show flux variability.
Radio-loud objects (blazars) show large amplitude flux variations 
relative to their
radio-quiet counterparts (Seyfert galaxies) in all the three bands, namely soft, hard
and total bands. Such large amplitude flux variations are expected if the 
X-ray emission in radio-loud sources are dominated by radiative processes
in their relativistic jets.

A total of 8 NLSy1 galaxies are in our sample. They are a separate class of
AGN with soft X-ray excess \citep{1996A&A...305...53B} showing 
fast large amplitude X-ray flux variations \citep{1995MNRAS.277L...5P,
1999ApJS..125..317L} and having FWHM  of the H$\beta$ line less than 2000 km $s^{-1}$ \citep{1985ApJ...297..166O}. Some of the peculiar properties of these sources have
been attributed to them having higher accretions rates compared to the
conventional broad line Seyfert galaxies \citep{2000ApJ...542..161P}. 
Considering all the NLSy1 galaxies studied here as a whole, from 
statistical tests, no difference is found in the variability behaviour
between soft and hard bands.
 However, three sources, namely 1H 0323$+$342, MCG$+$04$-$22$-$042 and PDS 456
 show high $F_{\rm{var}}$ values in the hard band 
relative to the soft band. MCG$+$04$-$22$-$042, 1H 0323$+$342 and PDS 456 are detected at 
1.4 GHz with flux densities of 10 mJy, 614 mJy and 22 mJy respectively 
\citep{1998AJ....115.1693C}. As the above three NLSy1 galaxies are detected in the radio band and as
they also show more amplitude of variability in the hard band relative to the
soft X-ray band, it is likely
that these three sources have a significant hard X-ray contribution from 
relativistic jets. Of them, 1H 0323+342 is a gamma-ray emitting NLSy 1 
galaxy 
\citep{2009ApJ...707L.142A}, suggesting the presence of a relativistic jet 
in it similar 
to blazars. Removing these three sources, from the list of NLSy1 galaxies
and classifying the remaining sources as radio-quiet NLSy1 galaxies, 
we find from U test, that in radio-quiet NLSy1 galaxies variations in soft band
is more than the variations in hard band. However, KS test report of no
difference in  F$_{\rm{var}}$ values between hard and soft bands. 
In the soft X-ray band our statistical tests indicate that radio-quiet 
NLSy1 galaxies are more variable than Seyfert 1 galaxies with broad optical 
emission lines similar to that reported by \cite{1999ApJS..125..317L}.
 However, from both U and KS test it is evident that there is 
no difference in the F$_{\rm{var}}$ statistics
in the hard X-ray band between radio-quiet NLSy1 galaxies and their broad line counterparts.
This increased F$_{\rm{var}}$ in the soft band in radio-quiet NLSy1 galaxies might be due to them
having lower BH masses than broad line Seyfert galaxies in accordance to the 
negative correlation between F$_{\rm{var}}$ and BH mass (see section 4.3).
However, recently from spectro-polarimetric observations of one NLSy1 galaxy
PKS 2004$-$447, \cite{2016MNRAS.458L..69B} suggested that the now existing notion of NLSy1 galaxies having low BH masses need not be true and NLSy1 galaxies too have
BH masses similar to blazars. If it is indeed the case, then the 
larger F$_{\rm{var}}$ seen in NLSy1 galaxies compared to their broad line
counterparts might be because of them having X-ray flux variations caused
by physical processes other than Seyfert 1 galaxies.
 In radio-quiet objects, no statistically significant differences in 
flux variations between soft and hard bands is found both in 
Seyfert 1 and Seyfert 2 galaxies. Also, we find both Seyfert 1 and
Seyfert 2 galaxies showing similar $F_{\rm{var}}$ characteristics in all
the three bands, namely, soft, hard and the full energy bands. 
Though these results pertain to variations on short time scales, 
from long term variability studies in the hard X-ray band, Seyfert 2 galaxies 
are found to be marginally more variable compared to Seyfert 1 
galaxies \citep{2007A&A...475..827B, 2014A&A...563A..57S}.

\begin{table*}
\caption{\label{lagresults}Results of correlation analysis. Here, P and R are the probability of no correlation and the correlation coefficient 
respectively.}
\begin{tabular}{@{}llrrrrr@{}}
\hline
Name  & OBS ID & Slope  & Intercept & $\chi^2_{\rm{red}}$ & P & R  \\
\hline
Mrk 421       & 10002016001 &  0.005   $\pm$ 0.000  & 0.081  $\pm$ 0.003    & 0.900 & $<$ 10$^{-5}$  &  0.793  \\
             & 60002023022 &  0.003   $\pm$ 0.000  & 0.062  $\pm$ 0.006    & 3.030 & $<$ 10$^{-5}$  &  0.802  \\
             & 60002023025 &  0.001  $\pm$ 0.000  & 0.133  $\pm$ 0.000  & 6.954 & $<$ 10$^{-5}$  &  0.524  \\
             & 60002023027 &  0.000  $\pm$ 0.000  & 0.119  $\pm$ 0.000     & 1.984 & $<$ 10$^{-5}$   &  0.709  \\
             & 60002023031 &  0.000  $\pm$ 0.000  & 0.176  $\pm$ 0.000 & 2.924 & $<$ 10$^{-5}$  &  0.876  \\
             & 60002023035 &  0.001  $\pm$ 0.000  & 0.169  $\pm$ 0.000 & 3.562 & $<$ 10$^{-5}$ &  0.646  \\
3C 273        & 00015012001 &  0.085   $\pm$ 0.034    & $-$0.186 $\pm$ 0.262    & 1.193 & 0.010     &  0.735  \\
             & 00015016001 &  0.049   $\pm$ 0.032    & 0.168  $\pm$ 0.187    & 0.984 & 0.029     &  0.658  \\
             & 00015017001 & $-$0.085   $\pm$ 0.028    & 1.038  $\pm$ 0.208    & 1.809 & 0.015     & $-$0.678  \\
NGC 5548     & 60002044005 & $-$0.192   $\pm$ 0.027    & 1.060  $\pm$ 0.077    & 1.301 & $<$ 10$^{-5}$ & $-$0.537  \\
NGC 1365     & 60002046007 & $-$0.093   $\pm$ 0.007    & 0.679  $\pm$ 0.018    & 1.890 & $<$ 10$^{-5}$ & $-$0.718  \\
\hline
\end{tabular}
\end{table*}

\subsection{Spectral variability}
To have an idea of the spectral variations of the sources relative to their
brightness, correlation between HR and total flux has been studied. The
HR is computed  from light curves that cover a wide energy range. Therefore
the disadvantage in using HR to characterize spectral variations is that they
do not identify spectral components that are responsible for the observed
variations measured over a band, however, they are the simplest one
to study spectral variations in a model independent way.  The majority of
sources in our sample do not show any correlation between HR and flux 
variations. However, some sources do show correlations between flux and 
spectral variations.  In the radio-loud category, objects such as 
Mrk 421 and 3C 273 (most of the time), showed a spectral
hardening with increasing brightness level,  i.e, a hardening when brightening
trend is noticed. Among the radio-loud sources that showed a 
harder when brighter trend, Mrk 421 belong
to the HSP type, where the harder when brighter trend is often seen
\citep{1990ApJ...356..432G,1998ApJ...492L..17P,2016ApJ...819..156B}. In the
FSRQ 3C 273 (most of the FSRQs are LSP blazars), we find both 
harder when brighter and softer when brighter trend.  We do not as yet have an 
unambiguous knowledge of the physical parameters that cause spectral 
variations. However, the harder when brighter trend seen in the 
sources studied here, namely, Mrk 421 and 3C 273  might 
be due to the emergence of a hard X-ray tail produced in the relativistic 
jets in the
high brightness states in these objects. Though the overall X-ray spectrum of 
HSP blazars is dominated by the high energy tail of synchrotron emission
(contrary to LSP blazars which is dominated by IC emission) it has recently 
been noticed that in the HSP blazar Mrk 421, during in its 
low brightness state, 
excess emission is noticed above 20 keV which is attributed to the emergence 
of the IC emission \citep{2016ApJ...827...55K}.
This clearly indicates that phenomena of flux and spectral variability in 
Mrk 421 and other blazars is complex.
To identify the nature
of the hard X-ray emission in our sample of radio-loud {\it vis-a-vis} 
radio-quiet sources, we generate the 
NuSTATR hard X-ray spectra and fit them with a simple power law model. The results
of this spectral fitting for a BL Lac Mrk 421 and a Seyfert galaxy 
PKS 1409$-$651 is shown in 
Figure~\ref{spectra}. From the figure it is clear that the hard X-ray emission 
in Mrk 421 (and other blazars) is likely dominated either by jet
synchrotron or inverse Compton processes or a combination of both 
assuming a leptonic jet model. 
For Seyfert galaxies, power law model is not a good fit to the spectra, thus
indicating the hard X-ray emission as due to thermal Comptonization processes. 
The  hardening when brightening 
trend noticed here in the radio-loud objects 
Mrk 421 and 3C 273 has been reported earlier 
\citep{2000MNRAS.312..123M,  2007ApJ...656..691G, 2008ApJ...678...78G}. 
For two Seyfert 
galaxies, namely NGC 5548 and NGC 3516 a softening when brightening trend 
is noticed. 
Such softening when brightening trend is known for other Seyferts at low energies
below 1 keV \citep{2009MNRAS.399.1597S} as well as at energies above 2 keV \citep{2014A&A...563A..57S}.

\subsection{$F_{\rm{var}}$ v/s black hole mass}
We show in Figure~\ref{Fvar_correlation} the relation between $F_{\rm{var}}$ in the 3$-$79 keV energy
range and black hole (BH) mass. The BH masses are collected from the 
online BH mass data base at the Georgia State University 
\footnote{http://www.astro.gsu.edu/AGNmass/}, which is a compilation
of BH masses determined from spectroscopic monitoring observations 
\citep{2015PASP..127...67B}. Of the 81 objects in our sample, BH masses are
available for only 50 objects. Among those 50 objects, only 37 objects are
found to be variable and they are thus used
to study the correlation if any between $F_{\rm{var}}$ and BH mass.
Figure~\ref{Fvar_correlation} indicates that there is an
anti correlation between $F_{\rm{var}}$ and BH mass. Both Spearmann rank correlation
and Kendalls $\tau$ tests on the data indicate that the anti correlation is
significant with $>$ 99.9 percent. Using a simple linear least squares fit
to the data by including the error in $F_{\rm{var}}$ we find
\begin{equation}
F_{var}  =  -(0.047\pm0.009) \times \log M_{BH} + (0.457\pm0.072)
\end{equation}
Therefore, there is in general a trend for high mass objects to be less
variable though there is a large scatter in the $F_{\rm{var}}$  v/s M$_{BH}$ 
diagram. Studies of X-ray variations in the soft X-ray band on a large 
number of sources do show a negative correlation between $F_{\rm{var}}$ 
and BH mass similar to the one observed here 
\citep{1995MNRAS.273..923P, 2005MNRAS.358.1405O, 2012A&A...542A..83P}.

\subsection{$F_{\rm{var}}$ v/s 3$-$10 keV luminosity}
The relation between $F_{\rm{var}}$ and the luminosity of the source in the 
energy range 3$-$10 keV is shown in Figure~\ref{Fvar_correlation}. 
To determine the flux, we have fitted power law to the observed spectra
 for all the sources 
with fixed galactic column density $n_{H}=1.63 \times 10^{20} cm^{-2}$.
This was then used to calculate  
the luminosity in the 3 $-$ 10 keV range.
 From the figure there is a hint that
low luminosity objects are more variable than their high luminosity 
counterparts.

\subsection{Short time scale variability}
Short time scales of flux variations in AGN taken as the flux 
doubling/halving time scale in this work are more often seen at 
high energies in the $\gamma$-ray band,
for example, ~4 min in PKS 2155$-$304  \citep{2007ApJ...664L..71A}
and $<$ 15 min in Mrk 421 \citep{1996Natur.383..319G} and 
14 min in Mrk 421 in the NuSTAR band \citep{2015ApJ...811..143P}.  
A recent systematic search for flux doubling/halving time scale
less than 15 minutes in the 
soft X-ray (0.2$-$10 keV) band, from {\it Swift}/XRT observation
of blazars by \cite{2015ApJ...802...33P} has yielded a negative result.
Observations of short time
scale flux variations in the high energy band of the electromagnetic 
spectrum is very important, as this can set an upper limit on 
the size of the emitting region (R$_s$) via the relation 
$R_s < \delta c \tau_{var}/(1 + z)$ and consequently, can help in constraining
the emission processes. Here, $\tau_{var}$ is the time scale of variability, 
$z$ is the redshift of the source and $\delta$ is the Doppler factor.
Doppler factor is defined as $\delta^{-1} = \gamma (1 - \beta \cos\theta)$, 
where, $\gamma$ is the Lorentz factor of the jet, $\beta = v/c$ is the 
ratio of the jet speed to the speed of light in vacuum and $\theta$ is the 
viewing angle, the angle between the jet and the observers line of sight.
In this work, by characterising the variability time scale by either the
flux doubling/halving timescale we found the shortest flux doubling/halving
time scale of 1.86 $\pm$ 1.02 min for the Seyfert 2 galaxy
NGC 1052. Using
this observed  $\tau_{var}$ we find $R_s < 3.3 \times 10^{13} (\delta/10)$ cm. 
Such short time scales of variations are also seen in few radio-loud 
sources. 
In our sample, we were able to derive flux doubling/halving time scale
for 16 sources, of which 11 sources have 
time scales less than 10 min.  Thus, this is the first report of the detection 
of statistically significant hard X-ray flux variability in a large number of 
AGN with flux doubling/halving time scale less than 10 minutes. The only 
other report of short time scale flux doubling/halving variability in 
the hard X-ray band available in the literature is for the source Mrk 421 
\citep{2015ApJ...811..143P}. On analysis of the blazar light curves
for which we were able to derive flux doubling time scales, we found 
that the flares are asymmetric in nature showing a quick rise and slow decay.
This
might be due to the difference between the particle acceleration and
synchrotron cooling time scales.

\section{Summary}
In this work we have examined 176 observations of AGN from 
{\it NuSTAR} to search for hard \mbox{X-ray} flux variations in them, 
characterize 
their variability and to see how the flux variations are related to 
other AGN properties. Key findings of this work are summarized below

\begin{enumerate}
\item A total of 81 sources (3 FSRQs, 4 BL Lac objects, 
24 Seyfert 1 galaxies, 42 Seyfert 2 galaxies and 8 NLSy1 galaxies) over 176 
sets of observations are studied for hard X-ray flux variability on hour time 
scales. We find evidence of X-ray flux 
variations in about 65\% of the sources in our sample.

\item NLSy1 galaxies are found to show the highest DC of variability
of about 87\% followed by BL Lac objects that show a DC of 82\%. 
Both Seyfert 1 and 2 galaxies show similar DC of about 56\%. 
The lowest DC of variations of about 23\% is shown by FSRQs. 

\item {In the 3 $-$ 79 keV band}, BL Lacs have a weighted mean  $F_{\rm{var}}$ of 
0.310 $\pm$ 0.138, while FSRQs have a weighted mean $F_{\rm{var}}$ of 0.078 $\pm$ 0.028. 
In the radio-quiet category, Seyfert 2 galaxies have a higher weighted mean 
$F_{\rm{var}}$ of 0.149 $\pm$ 0.108 relative to Seyfert 1 galaxies that have
a weighted mean  $F_{\rm{var}}$ of 0.092 $\pm$ 0.061. 
Both U and KS test reject the null hypothesis of differences in variability 
between FSRQs and BL Lac as well as Sy1 and Sy2 galaxies. 
In the soft X-ray band, radio-quiet NLSy1 galaxies are more variable than 
their broad line counterparts. Although, the BL Lacs and FSRQs
examined here  
 are more variable than the NLSy1 galaxies  in the
hard band, statistically in the soft bands their variations are 
indistinguishable.

\item Majority of the sources do not show any correlation between HR and 
count rate in the 3$-$79 keV energy range  suggesting no spectral variability. 
However, a small fraction of sources show significant correlation between 
HR and count rate in the 3$-$79 keV energy range. The BL Lac
object Mrk 421 showed a hardening 
when brightening trend and the FSRQ 3C 273 showed both hardening and 
softening when brightening behaviour. Among the three
epochs of observations, for which 3C 273 has shown variability, in two
epochs a hardening when brightening trend is observed, while in one epoch
a softening when brightening trend is noticed. In the radio-quiet
category, two Seyfert galaxies namely, NGC 5548 and NGC 3516 showed a 
softening when brightening behaviour. This is opposite to 
that shown by the BL Lac source Mrk 421. 

\item Sources hosted by  massive BHs are less variable on hour time scale than 
their less massive counterparts. Also, $F_{\rm{var}}$ values show a hint for 
a negative correlation with luminosity of the sources in the 3$-$10 keV band.

\item We are able to estimate 
flux doubling/halving time scale for a total of 16 sources in our sample,
of which 11 sources show flux doubling/halving time less than 10 minutes. 

\end{enumerate}

We caution on the small number of BL Lacs, FSRQs and NLSy1 galaxies
used in this work. 
Future observations of a large number of BL Lacs, FSRQs and NLSy1 galaxies are 
indeed needed to confirm the findings 
reported here.

\section*{Acknowledgments}
We thank the anonymous referee for his/her detailed comments that helped
to improve the manuscript.
The comments by Dr. Markus Bottcher on an initial version of the 
manuscript is thankfully acknowledged. 
This research has made use of data from the 
NuSTAR mission, a project led by the 
California Institute of Technology, managed by the Jet
Propulsion Laboratory and funded by
the National Aeronautics and Space Administration. 
We thank the {\it NuSTAR} Operations, Software and Calibration teams for 
support with the execution and analysis of these observations. 
This research has made use of the {\it NuSTAR} Data Analysis 
Software (NuSTARDAS) jointly developed by the ASI
Science Data Center (ASDC, Italy) and the 
California Institute of Technology(USA).

\section*{Note Added in Proof}
We thank Dr. Madsen of the NuSTAR operations team for bringing to our attention improved background filtering for South Atlantic Anomaly (SAA)  passage made available since the work in this paper as well as the ways to identify flares due to SAA from inspection of the background light curves. All data have now been checked for the existence of such events in our analysis and the updated results are given in this version of the manuscript.




\bibliographystyle{mnras}

\label{lastpage}
\end{document}